\documentclass[]{AIAA}

\usepackage{varioref}
 \usepackage{wrapfig}
 \usepackage{threeparttable}
 \usepackage{subfigure}
 \usepackage{subfigmat}
 \usepackage{lettrine}

\usepackage{graphics} 
\usepackage{epsfig} 
\usepackage{amsmath} 
\usepackage{amssymb}  
\usepackage{amscd} 
\usepackage{subfigure}
\usepackage{verbatim}

 \newcommand{\htext}[1]{}

\newcommand{\BB}{{\mathcal  B}}

\newcommand{\CC}{\mathcal{C}}
\newcommand{\WW}{\mathcal{W}}

\newcommand{\TT}{\mathcal{T}}
\newcommand{\FF}{\mathcal{F}}
\newcommand{\rv}[1]{\mathbf{#1}}

\newcommand{\Rset}{{\mathbb R}}

\newcommand{\Nset}{{\mathbb N}}

\newcommand{\norm}[1]{\left\lVert#1\right\rVert}

 \long\def\symbolfootnote[#1]#2{\begingroup%
\def\thefootnote{\fnsymbol{footnote}}\footnote[#1]{#2}\endgroup}

\begin{document}
\title{\LARGE \bf
Aircraft Proximity Maps\\
Based on Data-Driven Flow Modeling}

\author{Erwan Sala\"{u}n\footnote{Postdoctoral Fellow, School of Aerospace Engineering, 270 Ferst Drive, {\tt erwan.salaun@gatech.edu}.},}
\affiliation{Georgia Institute of Technology, Atlanta, GA, 30332}
\author{Maxime Gariel\footnote{Postdoctoral Associate, Laboratory for Information and Decision Systems, 77 Massachusetts Avenue, {\tt mgariel@mit.edu}. The work was realized when Maxime was a PhD candidate in the School of Aerospace Engineering of Georgia Institute of Technology.},}
\affiliation{Massachusetts Institute of Technology, Cambridge, MA, 02139}
\author{Adan E. Vela\footnote{Ph.D. Candidate, School of Mechanical Engineering, 801 Ferst Drive, {\tt aevela@gatech.edu}.},
and Eric Feron\footnote{Professor of Aeronautics and Astronautics, School of Aerospace Engineering, 270 Ferst Drive, {\tt feron@gatech.edu}}
}
\affiliation{Georgia Institute of Technology, Atlanta, GA., 30332}



 \begin{abstract}
With the forecast increase in air traffic demand over the next decades, it is imperative to develop tools to provide traffic flow managers with the information required to support decision making.  In particular, decision-support tools for traffic flow management should aid in  limiting controller workload and complexity, while supporting increases in air traffic throughput. Indeed, the growth of air transportation is conditioned on the ability to maintain acceptable safety. While many decision-support tools exist for short-term traffic planning, few have addressed the strategic needs for medium- and long-term planning for time horizons greater than 30 minutes.  This paper seeks to address this gap through the introduction of 3D aircraft proximity maps that evaluate the future probability of presence of at least one or two aircraft at any given point of the airspace.  Three types of proximity maps are presented: (i) presence maps that indicate the local density of traffic, (ii) conflict maps that determine locations of potential conflicts and their corresponding probabilities, and (iii) outlier proximity maps that show the probability of conflict due to aircraft not belonging to dominant traffic patterns.  These maps provide traffic flow managers with information relating to the complexity and difficulty of managing an airspace.   The intended purpose of the maps is to anticipate how aircraft flows interact, and how outliers impact the dominant traffic flow for a given time period. This formulation is able to predict which ``critical'' regions may be subject to conflicts between aircraft, thereby requiring careful monitoring and additional effort to manage the airspace.  These probabilities are computed using a generative aircraft flow model.  Time-varying flow characteristics, such as geometrical configuration, speed,  and probability density function of aircraft spatial distribution within the flow, are determined from archived Enhanced Traffic Management System data, using a tailored clustering algorithm.  Aircraft not belonging to flows are identified as outliers.
\end{abstract} 
\maketitle


 \section{Introduction}
\label{intro}
ACCORDING to the predicted air traffic growth, current airspace capacity limits, i.e. the maximum number of aircraft allowed in an airspace, will be increasingly exceeded throughout the National Airspace System (NAS) in the coming decades~\cite{nextgen:2009}.  Currently, the capacity of an airspace is measured according to the peak number of aircraft allowed in the considered sector within any one minute period~\cite{etms_02}. When the predetermined thresholds are exceeded, congestion alerts are triggered. In practice, capacity thresholds are set by the ability of controllers to handle traffic and to maintain manageable workloads for worst-case planning over a wide range of traffic patterns~\cite{majumdar2005route}.  Hence, current airspace capacity thresholds are independent of any changes in the traffic configurations that may occur due to weather or other extraneous traffic perturbations.  Such a ill-defined method for establishing and measuring airspace capacity often leads to under-utilization of the airspace under nominal conditions.  Or conversely, when convective weather is present, over-utilization is not uncommon.  To accommodate high levels of air traffic throughput and to effectively manage airspace utilization while maintaining safety, intrinsic descriptors of traffic and the airspace are required to support airspace management in the redesign or adjustment of flow configurations. These measures are often referred to as ``complexity measures'', since they aim at providing an image of the difficulty or complexity of managing the considered airspace and traffic configuration.

There has been a significant volume of research related to the estimation of air traffic control complexity, and its application to  determine airspace capacity. One method for defining and establishing air traffic control complexity is dynamic density~\cite{laudeman:1998,sridhar:1998}. This metric is a weighted sum of the local density of aircraft; the number of heading, altitude, and speed changes; and a variety of other factors. The weighting factors are determined and validated through human-in-the-loop experiments. Dynamic density makes use of instantaneous aircraft configurations, and therefore requires accurate prediction of traffic to hold any significant forecasting value. The impact of uncertainties on aircraft trajectories on predicting dynamic density has been studied in~\cite{sridhar:2009,sridhar:1998}, and was demonstrated to cause short-comings in forecasting complexity.  Another intrinsic measure of air traffic control complexity, proposed in~\cite{lee_delahaye:gnc09}, analyzes the stability of a nonlinear dynamical system corresponding to the considered traffic pattern. This analysis method is attractive for its ability to seamlessly account for aircraft position uncertainties.  An alternative approach toward managing complexity is presented in~\cite{yousefi2004}.  The authors propose a dynamic sectorization of the airspace based on air traffic controller workload, which leads to the generation of another kind of complexity map.

Recently, significant effort has been placed into understanding the probability of conflict and the associated resolution-commands required to ensure separation. For example, several tools are presented in~\cite{PaielliErzberger:1997,Irvine:2002,blom:ecc_03} to determine the probability of conflict for a given geometric traffic configuration, with and without aircraft position uncertainty. In~\cite{lee:jgcd_09}, an ``input-output'' approach is proposed that generates complexity maps based on the conflict-resolution commands required to accommodate disturbances such as the entrance of an exogenous aircraft into the airspace. In~\cite{prandiniCSP:2009}, the authors use aircraft position and intent information to compute the probability of presence of the aircraft at any point in the airspace within a given time frame. In~\cite{krozel2007automated,krozel2007maximum}, the authors propose a complexity metric that is a function of the local average velocity vector, the density of aircraft and the proximity of an aircraft to the considered point of the airspace, in order to evaluate the solutions computed by their traffic flow management (TFM) algorithm.

Nevertheless, a key piece of information forming the foundation of the previously cited works is knowledge of the initial aircraft positions and trajectories (``known'' with or without uncertainty). Such information, while useful for short-term air traffic management, e.g. planning a safe trajectory for an intruder aircraft through the airspace (e.g.~\cite{prandiniCSP:2009}), is not well-suited for mid-term or long-term air traffic management, that is, for time horizons greater than 30 minutes. Indeed, from a Traffic Flow Management perspective, strategic planning requires knowledge of air traffic patterns and flows characteristics (e.g. geometrical configuration, flight plans, distances between two consecutive aircraft), and not the current position and intent information of individual aircraft. As explained in~\cite{song_2006}, considering aircraft flows enables the computation of more predictable and perturbation-resistant estimates than considering individual aircraft.  Representing air traffic as flows provides a more meaningful and higher-level view of air traffic to traffic flow managers.  In fact, such a representation is also consistent with an air traffic controllers understanding of a sector.  As explained in \cite{histon2002sca, histon2002introducing}, one such way air traffic controllers abstract sectors is according to dominant traffic flow patterns and traffic flow interactions.  Few works have developed TFM tools with this viewpoint in mind.  The authors of~\cite{song_2006} provide a list of flow features to describe flow pattern, such as the number of flows, the major flow and its size, the number of crossing flows, etc. In~\cite{song_2009}, these criteria allow the authors to estimate and predict sector capacity as a function of the traffic flow pattern, and to study the impact of severe weather. While the list of flow features established in~\cite{song_2006} provides a notion of the traffic complexity (given as a scalar number), more information may be needed by traffic flow managers to identify regions of high complexity. In~\cite{Salaun_DASC:2009}, a simplified approach to generate complexity maps, taking into account the influence of automated an conflict-resolution algorithm and the flows characteristics, is proposed. However, only two flows are considered, and the extension of the results of~\cite{Salaun_DASC:2009} to a realistic airspace (i.e. multiple intersections with multiple flows) is not obvious and may be computationally expensive.

Building upon the preliminary work in~\cite{salaun:gnc10}, this paper proposes a new kind of air traffic complexity maps aimed at supporting traffic flow management. These maps, referred to as ``(aircraft) proximity maps'', provide an estimation of probability of proximity of one or more aircraft to a region of airspace. For simplicity, the expression \emph{probability of proximity} is used throughout the paper instead of the correct (and more rigorous) expression \emph{probability an aircraft is within a given neighborhood of a point $P$ in the airspace}. Three kinds of proximity maps are developed, each with a different probability metric.  They are as follows:
\begin{itemize}
  \item \emph{Presence maps} depict the probability of presence of \emph{at least one} aircraft from a flow in a given volume of the airspace. These maps enable the identification of the main flows of aircraft and the regions of high aircraft density.
  \item \emph{Conflict maps} provide the probability of presence of \emph{at least two} aircraft from two different flows in a given volume of the airspace. These maps highlight regions of airspace with high probability of conflict or close proximity between aircraft, thereby indicating regions of the airspace that require additional attention from the air traffic controllers.
  \item \emph{Outlier proximity maps} give the probability of presence of an aircraft from the dominant traffic and an \emph{outlier aircraft} (defined in Section~\ref{sec:outliers}) in a considered volume of the airspace. These maps aid traffic flow managers in evaluating the impact of aircraft outliers on the dominant air traffic flows.
\end{itemize}
Regions of the maps with high values for the probability of presence, conflict, or outlier proximity are interpreted to be ``critical'' regions that require monitoring.

\begin{figure}
 \centerline{
 \includegraphics[width=16cm]{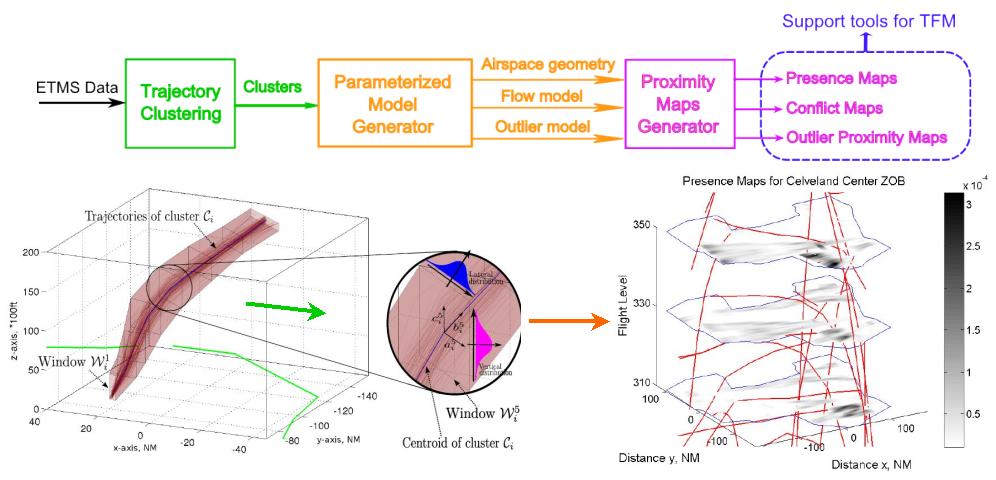}}
 \caption{Work-process flow from historical data to the generation of proximity maps.}\label{proposed_tools}
\end{figure}

The proposed proximity maps take into account flow characteristics in a manner consistent with TFM (i.e. aggregate flow descriptions such as average miles-in-trail), thereby making them an intuitive and effective strategic decision-support tool.  To enable the creation of the presence and proximity maps, a consistence model of airspace is required.  However, because a describing airspace is quite complex, we introduce a dimensionally-reduced model.   The dimensionally-reduced model serves as a generative model that is consistent with the tools we are proposing.  That is, the initial modeling provides a framework in which a parametric model can be established.  For this purpose, the clustering algorithm introduced in \cite{gariel10trajectory} is used to process archived Enhanced Traffic Management System (ETMS) data to generate a realistic air traffic flow model, i.e. the generative model.  The resulting model accounts for the aircraft spatial distributions and aircraft arrival distributions for each flow. Ultimately, the methodology for generating the traffic model provides a data-driven framework on which proximity maps are based. The process, starting from archived data to the generation of proximity maps, is illustrated in Fig.~\ref{proposed_tools}, with an example of clustered trajectories and presence maps.

The usage of proximity maps extends beyond displaying critical information.  They may be used for decision making in traffic flow management.  Examples include dynamically adjusting flow characteristics; redefining or creating new flight routes; or visualizing the influence of rogue aircraft (similar to~\cite{lee:jgcd_09}) or severe weather (as in~\cite{krozel2007maximum}) located in a specific airspace region. The tools presented in this article may also be utilized as decision support tools in conjunction with the Flow-Based Route Planning method used in~\cite{krozel2007automated,krozel2007maximum}, or other such tools. 

The remainder of the paper is organized as follows. Section~\ref{sec:data} presents a methodology for developing a traffic model through the extraction of air traffic flows and outliers from archived flight data.  The resulting model serves as a parametric generative model for computation of the proximity maps.  In the next section, an empirical analysis of the clustering is completed.  The analysis is completed to support assumptions and simplifying calculations used in the proximity maps. Section~\ref{sec:proximity} introduces the notion of the ``residual distance'' of an aircraft and describes how presence, conflict and outlier proximity maps are generated from the flow and outlier models. In Section~\ref{sec:example}, example proximity maps are generated for a representative region of airspace.  Finally, in Section~\ref{sec:conclusion}, our conclusions are presented.

\section{Airspace Modeling: Transformation of flight data to flow and outlier models}
\label{sec:data}
This section introduces a methodology to build mathematical models of aircraft operations within an airspace derived from ETMS data. The resulting model serves as a generative parametric model to create the proposed proximity maps.  To precede, useful definitions are first introduced. Next, the ETMS data set is described, followed by a brief discussion of the clustering algorithm used to generate the traffic models.  The clustering algorithm is then applied to an typical airspace, and analysis of the resulting traffic flow and outlier model is provided.

\subsection{Definitions}\label{sec:def}
Consider a 3D airspace $\mathcal{A}\in \Rset^3$, whose arbitrarily shaped boundaries are fixed and independent of the altitude. The coordinates of a point $P=(x,y,z)$ of the airspace are expressed in an East-North-Up frame $\mathcal{R}_0$, where the $x,y$ axes are expressed in nautical miles (NM) and the altitude $z$ is expressed in feet (ft); a simple latitude-longitude to flat earth projection is used, with the origin of the airspace is located at the center of the airspace.  For the scope and scale of the airspace (approximately 400 NM in the largest dimension), the error introduced by this approximation is insignificant in relation to the topological descriptors need to understand generate the proximity maps.  As customary, Flight Levels (FL) are used for the $z$-coordinate. Flight levels are discretized at 100 ft increment, such that FL 320 corresponds to an altitude of 32,000 ft.

\begin{figure}[htbp]
\centering
\includegraphics[width=0.9\textwidth]{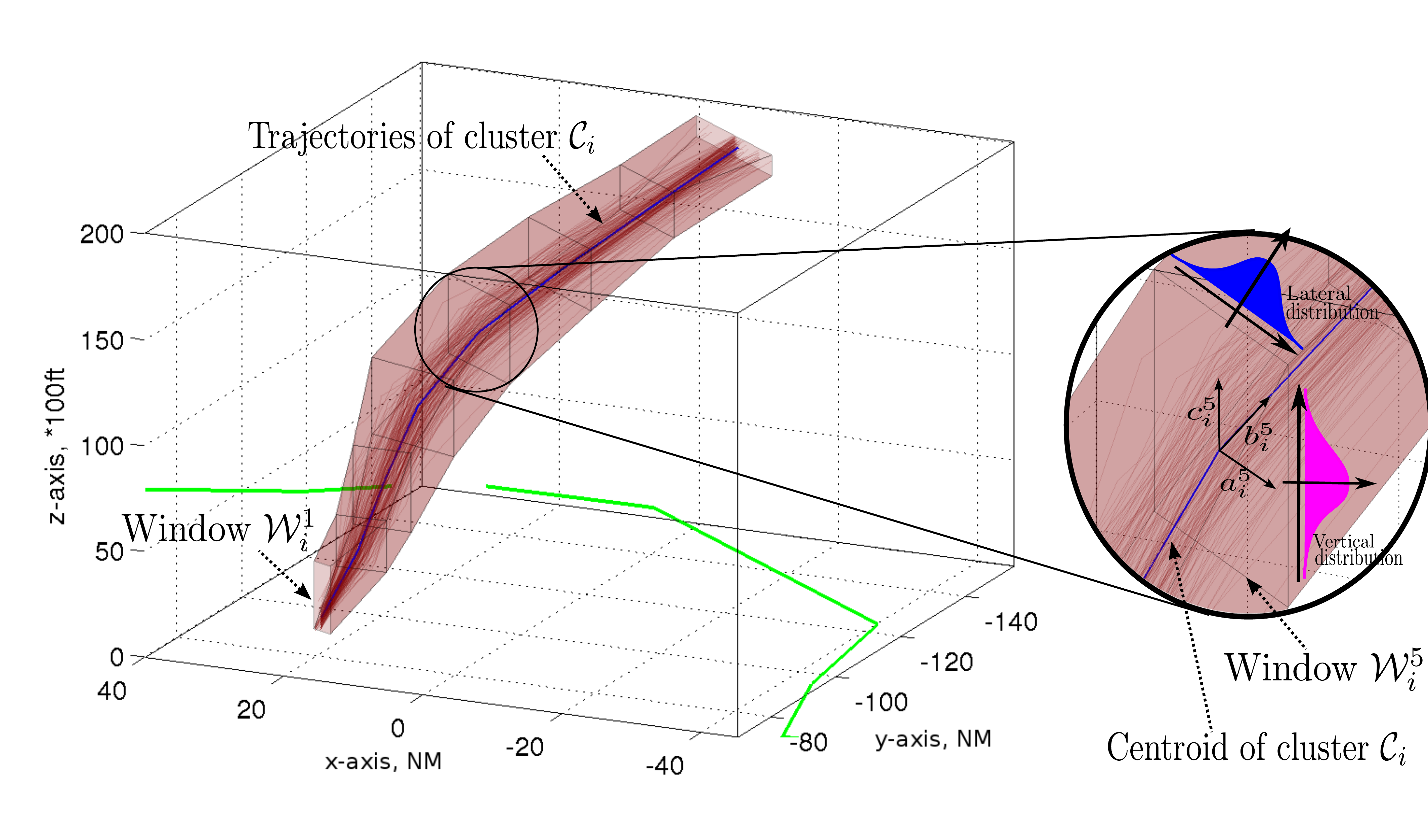}
\caption{A 3D representation of an ascending flow with its corresponding tube.}\label{fig:oneFlowWithTube}
\end{figure}

The following terms, used throughout the article, some of which are depicted in Fig.~\ref{fig:oneFlowWithTube}, are defined as follows:
\begin{itemize}

\item A \textit{trajectory} is a sequence of $l$ points defined by a set of 3D coordinates. All trajectories are resampled to have same number of points, $l = 8$. Trajectories are spatially sampled, instead of temporally sampled, in order to allow for identification of spatially similar trajectories. 

\item A \textit{cluster}, $\CC_i$, is a labeling of $m_i$ trajectories that are identified as ``similar'' by the clustering algorithm.

\item A \textit{track} $\TT_i$ is the centroid of the cluster $\CC_i$, i.e the trajectory ``center of mass'' of all the trajectories in the cluster. The track $\TT_i$ can be interpreted as the representative trajectory for cluster $\CC_i$.

\item The \textit{window} $\WW_i^k$ of cluster $\CC_i$ is located at the $k^{\text{th}}$ point along the associated track $\TT_i$.  The window $\WW_i^k$ is vertically orientated with its face aligned according to the heading of $\TT_i$ at the $k^{\text{th}}$ point towards the $(k+1)^{\text{th}}$ point.  The dimensions of each window are determined by the extrema of the $k^{\text{th}}$ points corresponding to each trajectory within the cluster labeling.

\item At each window $W_i^k$, the \textit{local coordinate frame} ($\overrightarrow{a_i^k},\overrightarrow{b_i^k},\overrightarrow{c_i^k}$) located at point $k^{\text{th}}$ point on track $\TT_i^k$.  The vertical axis, $\overrightarrow{c_i^k}$, is aligned with the global $z$ coordinate axis, while $\overrightarrow{a_i^k}$ align with the track $\TT_i$, such that it points towards $W_i^{k+1}$.

%
%

\item A \textit{box} $\BB_i^k, k=1\ldots l-1$, is the polyhedron defined by the convex hull of two consecutive windows, $\WW_i^k$ and $\WW_i^{k+1}$.

\item A \textit{tube} is the succession of all the boxes of a given cluster.

\item For every window $\WW_i^k$, the sampled \textit{lateral distribution of aircraft}, $f_{a_i^k}$, is computed from the orthogonal projections of the $k^{\text{th}}$ point of the trajectories in $\CC_i$ onto the $a_i^k$ axis of window $\WW_i^k$.

\item For every window $\WW_i^k$, the sampled \textit{vertical distribution of aircraft}, $f_{c_i^k}$  is computed from the altitudes of the $k^{\text{th}}$ point of the trajectories in $\CC_i$ along the $c_i^k$ axis.

\item The \textit{aircraft arrival rate parameter} $\lambda(t)$, is the expected number of aircraft entering the airspace region during a given time period. The arrival rate for each cluster $\CC_i$, $\lambda_i(t)$, is defined similarly. For the remainder of the paper, the considered time period is 15 minutes long.

\item A \textit{flow} $\FF_i$ is a 3-dimensional tube that contains the trajectories of cluster $\CC_i$. The tube is defined by a succession of $l$ windows  $\WW_i^k, k=1\ldots l$. Associated with a flow are its aircraft speed distribution, aircraft arrival rate, and its aircraft lateral and vertical distributions.
\end{itemize}

\subsection{ETMS Data Set}
The data used to construct the airspace traffic model is taken from the Enhanced Traffic Management System (ETMS) data set. Cleveland center is selected for the study because of its significance to the NAS. Indeed, a large share of the traffic to and from the North-East of the United States flies through the center. Furthermore, as noted in \cite{grabbe2004modeling}, Cleveland sector is ``one of the most congested and delay-prone Centers in the continental United States,'' and centrally to located to a number of high traffic regions.

\begin{figure}[htbp]
\centering
\includegraphics[width=0.7\textwidth]{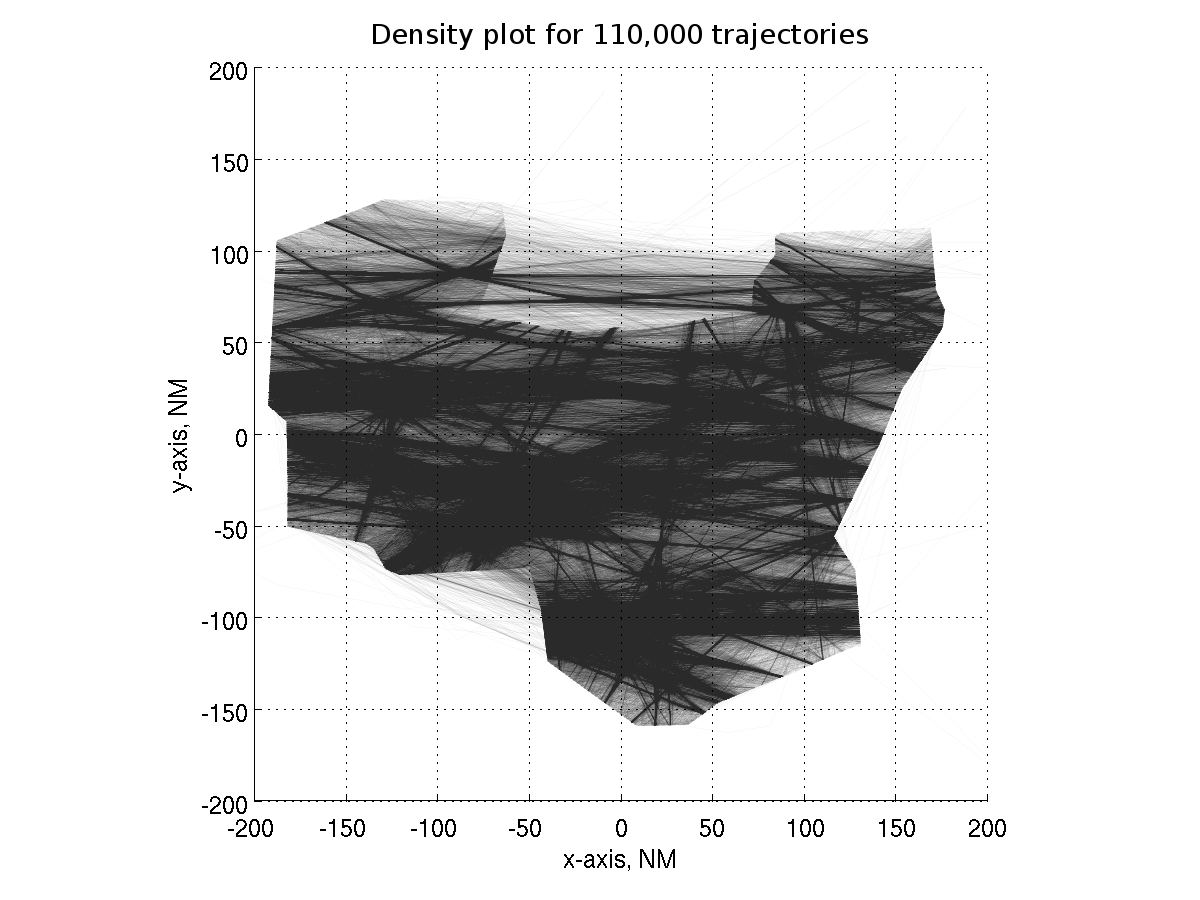}\caption{Density plot of a subset of the ``clean'' trajectories.}\label{fig:density}
\end{figure}

The ETMS data includes aircraft radar tracks (longitude, latitude, altitude) taken at approximately 1 minute intervals. During the 123 days (May to August 2005) covered by the data, all 526,840 aircraft trajectories with at least one point over FL 250 are considered; this subset of the data covers the majority of en route aircraft.  The dates 08/04/2005, 08/24/2005, and 08/26/2005 are not included in the clustering analysis, as the corresponding data was not available. ETMS data is notorious for its inconsistencies in altitudes, therefore a simple filtering algorithm on reported altitudes was used. After this initial filtering process, aircraft with inconsistencies in reported positions and altitudes are excluded. Flight data is deemed to be inconsistent if it exhibits physically unrealizable trajectories, e.g. rapid changes in altitude or position.  After filtering out inconsistent trajectories, a ``clean'' data-set of 338,060 trajectories remains. Statistics not requiring the altitude or the position information, such as the aircraft arrival rate into the airspace are computed on the entire en route data set, while the clustering of trajectories is run on the ``clean'' subset. Figure~\ref{fig:density} presents a randomly sampled density plot of 110,000 ``clean'' trajectories over Cleveland center (only 110,000 flights are plotted because of memory limitations). 

Major air traffic flows are clearly identifiable as the darkest regions of the airspace; many corresponding to defined jetways. The appearance of the gray shades indicates that some aircraft fly outside of the major flows.  The clustering algorithm presented in the next subsection is used to identify the major flows, thereby establishing to dominant traffic pattern.

\subsection{Trajectory Clustering}\label{sec:clustering}
The 338,060 trajectories are clustered using a slightly modified version of the algorithm presented in~\cite{gariel10trajectory}. The algorithm is defined according to the following steps:
\begin{enumerate}

\item  Clean and format the trajectories.  Inconsistent trajectories are excluded.  The remaining trajectories are spatially resampled according to a linear interpolation with a fixed number of equally space points. The linear approximation is sufficient as most aircraft fly along straight trajectories in this center.

\item  Augment the dimensionality of the trajectory data by including additional features such as heading, polar coordinates, etc.

\item Apply hierarchical clustering. Organize and separate the trajectories by altitude and attitude to create separate data sets. Level trajectories are split by flight level; climbing trajectories are sorted by destination (or airspace region exit) flight level; and descending trajectories are split by initial flight level.

\item Normalize each feature and concatenate the data into a single row vector for each flight.  Each hierarchical grouping - defined by altitude and attitude - has a corresponding matrix of data, including interpolated trajectory points and augmented data. Each row of the matrices corresponds to a flight, and each column corresponds to a feature.

\item Apply principal component analysis \cite{shlensTutorialPCA} on the previously described matrices and keep the projections onto the first 5 principal components, thereby reducing the dimensionality of the data.

\item Cluster the values of the projections using a density-based clustering algorithm (DBSCAN~\cite{ester1996density}).

\item Obtain clusters of trajectories and outliers for each altitude and attitude category. Outliers are trajectories that are not assigned to a cluster.
\end{enumerate}

Here, a key extension to the work presented in~\cite{gariel10trajectory} is to split the dataset into subsets, by altitude and attitude (climb, descent, level flight), to greater fidelity in modeling. This initial categorization results in 136 subsets. As noted, the trajectories for each category are clustered using the presented algorithm. The number of clusters and intra-cluster similarity can be adjusted by tuning input parameters to DBSCAN (the minimum number of points required to start a cluster, and the minimum distance between those points). With such a large sample set of trajectories, a two-pass approach is taken towards the clustering.  The first round identifies both large and small clusters.  A second application of DBSCAN is applied to large clusters with more discriminating parameters.

\begin{figure}
\centering
\subfigure[Top view]{\includegraphics{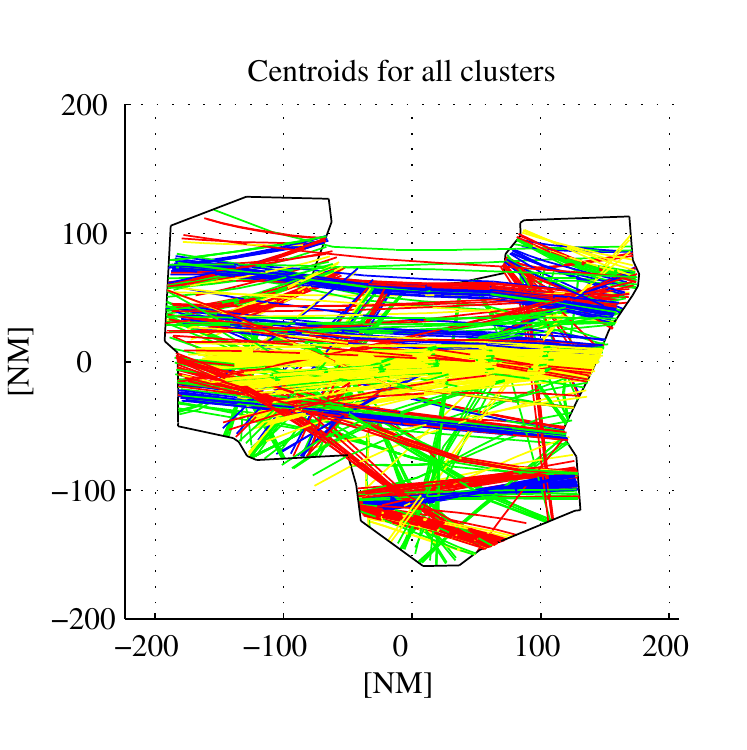}\label{fig:2dCentroids}}
\subfigure[Oblique view]{\includegraphics{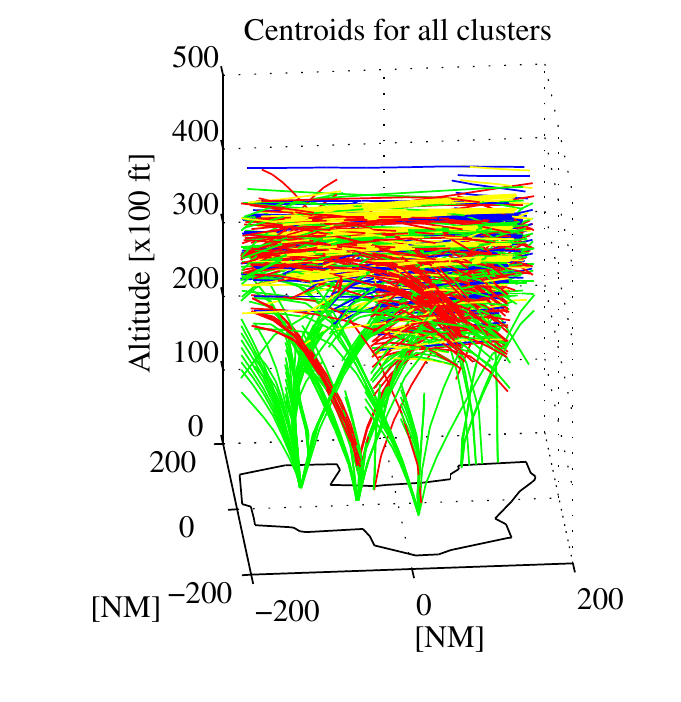}\label{fig:3dCentroids}}
\caption{Centroids for all traffic flow clusters.}\label{fig:ZTL39densityCentroids}
\end{figure}

Following categorization and clustering, approximatively 80\% of the trajectories are grouped into $685$ clusters, or flows.   Figure~\ref{fig:ZTL39densityCentroids} presents a 2D and 3D view of the centroids of the clusters. Blue lines represent westbound level traffic, yellow lines represent eastbound level traffic, green lines represent descending traffic, and red lines represent ascending traffic.  Eastbound traffic (yellow lines) appear to be solely passing through the center of the center. The reason is that Eastbound traffic is widespread over the entire center and no clear major flow is identifiable. Therefore, for each flight level, most of the Eastbound trajectories belong to the same cluster which is very wide.
In Figure \ref{fig:3dCentroids}, major airports of Cleveland center -- Cleveland Hopkins airport (CLE), Detroit Metropolitan Wayne County Airport (DTW) and Pittsburgh International Airport (PIT) -- are clearly identifiable by the clusters corresponding to descending air traffic. Given the greater variability in takeoff trajectories, in comparison to landing trajectories, few ascending clusters from airports are identified.  Primary reasons for variability include the diverse range in aircraft performance and pilot decisions characteristics.
%

\subsection{Clustering Analysis}
Applying clustering to the flight data identifies numerous flows over a broad range of trajectory types.  To better understand the structure of air traffic in relation to the clustering algorithm, results are analyzed over a broad range of parameters to determine when flow models of the airspace may lose validity.  Of particular interest is the proportion of outlier aircraft, that is, aircraft trajectories that are not assigned a cluster labeling.

\begin{figure}
\centering
\includegraphics{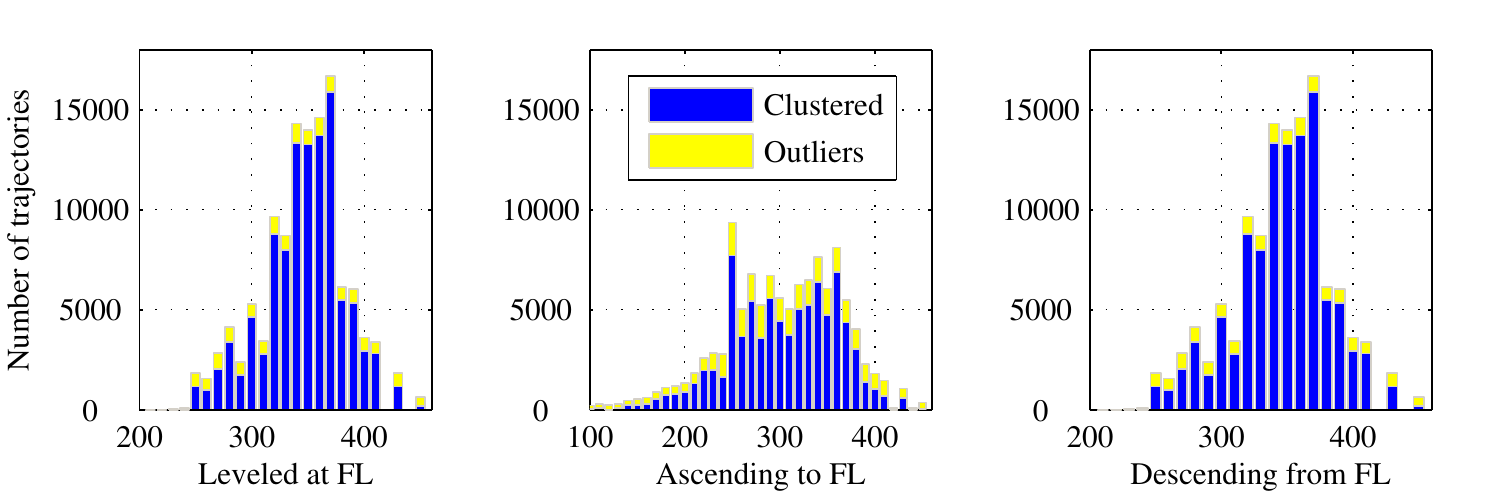}\caption{Distribution of clustered nad outlier flights by flight level.}\label{fig:outliersFL}
\end{figure}

Figure~\ref{fig:outliersFL} presents the distribution of clustered and outlier trajectories by flight level. ``Leveled at FL'' indicates that the altitude is constant for the corresponding aircraft trajectory; ``Ascending to FL'' indicates the destination flight level for the aircraft trajectory within the center; and ``Descending from FL'' indicates the initial flight level of the aircraft before starting its descent.  For the considered flight levels over 80\% of trajectories belong to a cluster. This result supports the hypothesis that flows exist, and can be meaningful towards describing air traffic.

\begin{figure}
 \includegraphics{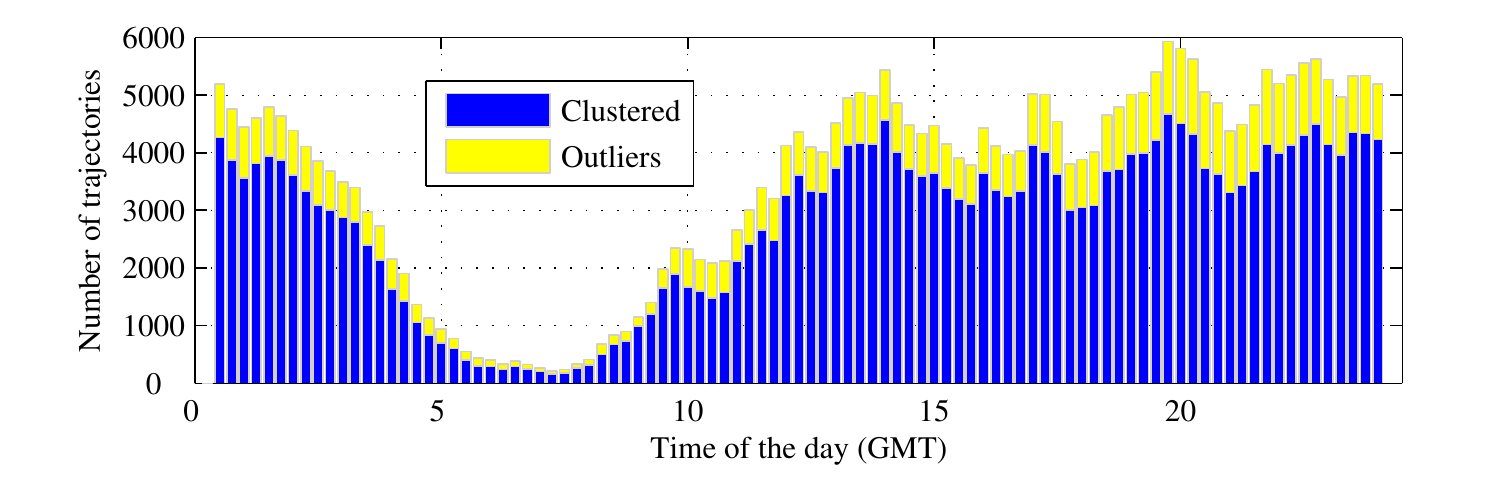}
\caption{Distribution of clustered and outlier flights by time of the day.}\label{fig:outliersTimeDay}
\end{figure}

\begin{figure}
\includegraphics{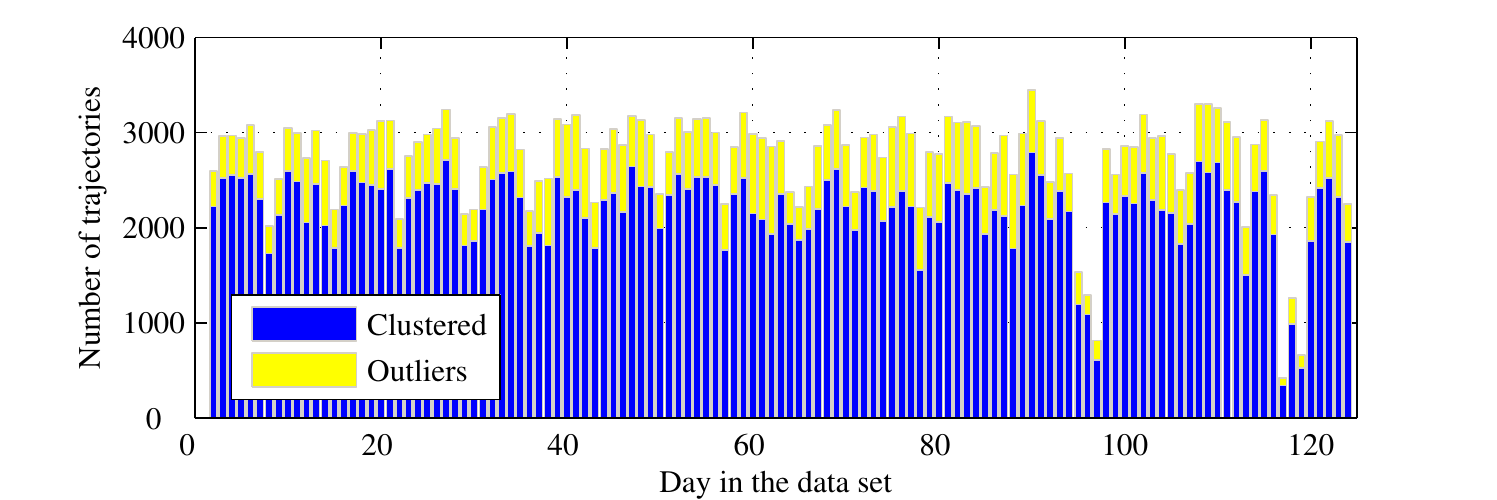}
\caption{Analysis of clustered and outliers flights by day of traffic.}\label{fig:clusteredTimeOfTheYear}
\end{figure}

Further analysis is completed to determine the fraction of trajectories clustered according to time of day, and date.  Figure~\ref{fig:outliersTimeDay} illustrates the distribution of clustered and outlier trajectories by time of the day. According to the figure, the fraction of outliers is relatively constant.  Meanwhile, the number of trajectories clustered varies throughout the day, with the maximum number of clusters being established during peak hours, i.e. between 11 a.m. and 2 a.m. GMT (5 a.m. and 8 p.m. local time). Figure~\ref{fig:clusteredTimeOfTheYear} presents the number of trajectories identified as part of a cluster, or identified as an outlier for each of the 123 days (1 = May 1, 2005).

The results presented here demonstrate that the clustering remains consistent over a broad range of parameterizations (time of day, date, altitude/attitude).  Therefore, clustering yields a model that can be utilized for subsequent complexity analysis.

%

\subsection{Outliers}
\label{sec:outliers}
While the vast majority of the traffic (approximatively 80\%) are clustered into flows, 20\% are designated to be outliers. Outliers are trajectories that are deemed ``too different'' from trajectories assigned flow labels by the clustering algorithm. ``Too different'' is a relative measure that is reflected by parameters of the clustering algorithm.  In order to approach a complete model of the airspace, it is vital that non-clustered trajectories are accounted for.  Outlier aircraft are particularly important to assessing air traffic control complexity because they act as perturbations to an otherwise static system.  As such, complete traffic models require inclusion of outlier aircraft that are not clustered according to standard traffic flows. The outlier model is given in the next section, with the complete parameterized air traffic model.

 \section{Creating a Parameterized Generative Model}
\label{sec:gen_model}
The previous section provided a methodology for clustering aircraft trajectories into flows and for determining which aircraft can be considered as outlier (non-clustered aircraft).  Using the clusters as a basis, the model is further analyzed to create a parameterized generative model.  Conceptually, the parameterized generative model provides an input to generate proximity complexity maps.  Parameterizing the airspace traffic model in this fashion allows for tools to test `what-if' scenarios or to create proximity maps without re-clustering and analyzing the air traffic data.  To parameterize traffic, the aircraft arrival process and the speed distribution for each flow are analyzed and modeled, as well as the outliers trajectories.

\subsection{Spatial Flow Distribution}\label{sec:ex_flow}
The trajectories labeled to a particular flow by the clustering algorithm do not follow a singular trajectory, hence the introduction of trajectory windows and flow tubes.  In fact, some flows have a large width, up to several hundred miles for Eastbound flight levels with a large throughput.

\begin{figure}[htbp]
\begin{minipage}{0.48\textwidth}
 \centering
\subfigure[Example flow]{\label{fig:exampleflow}\includegraphics{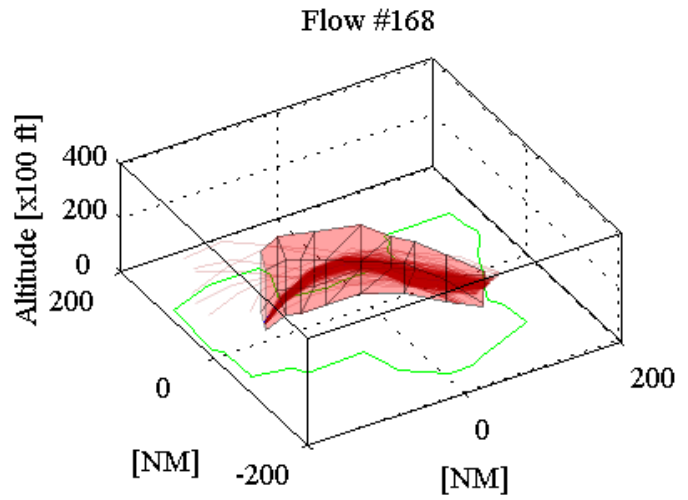}}
\subfigure[Distribution analysis at a window]{\includegraphics[width=0.99\textwidth]{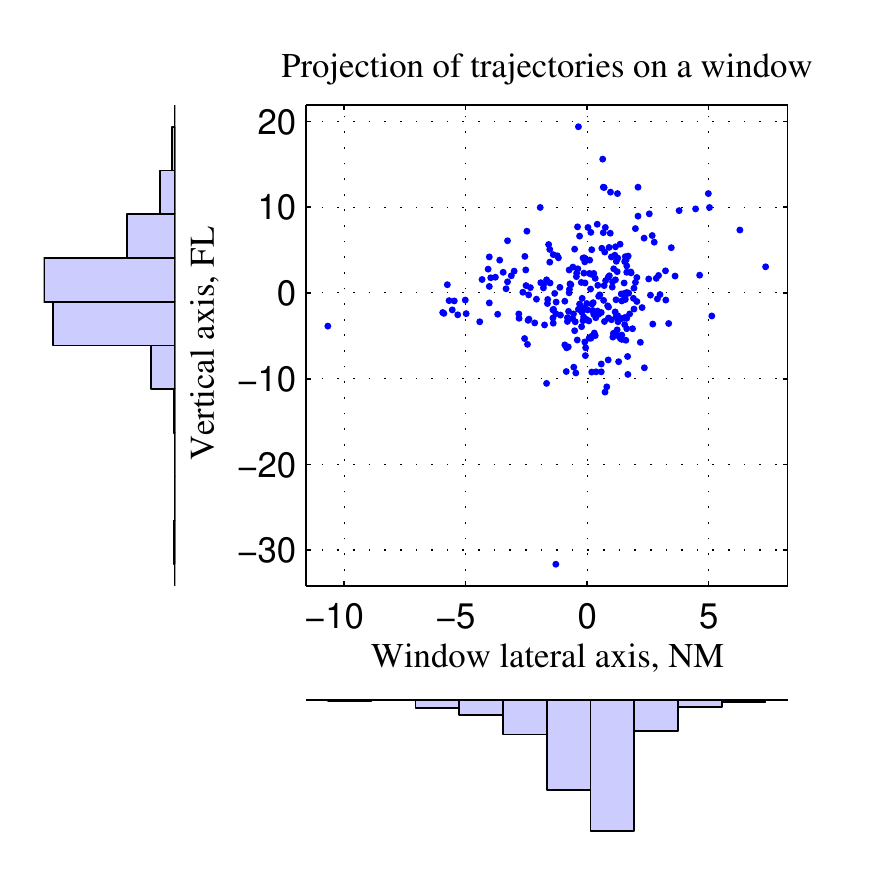}}
\end{minipage}
\begin{minipage}{0.49\textwidth}
 \centering
\subfigure[Distribution analysis for a flow]{\label{fig:flow_distribution}\includegraphics[width=0.9\textwidth]{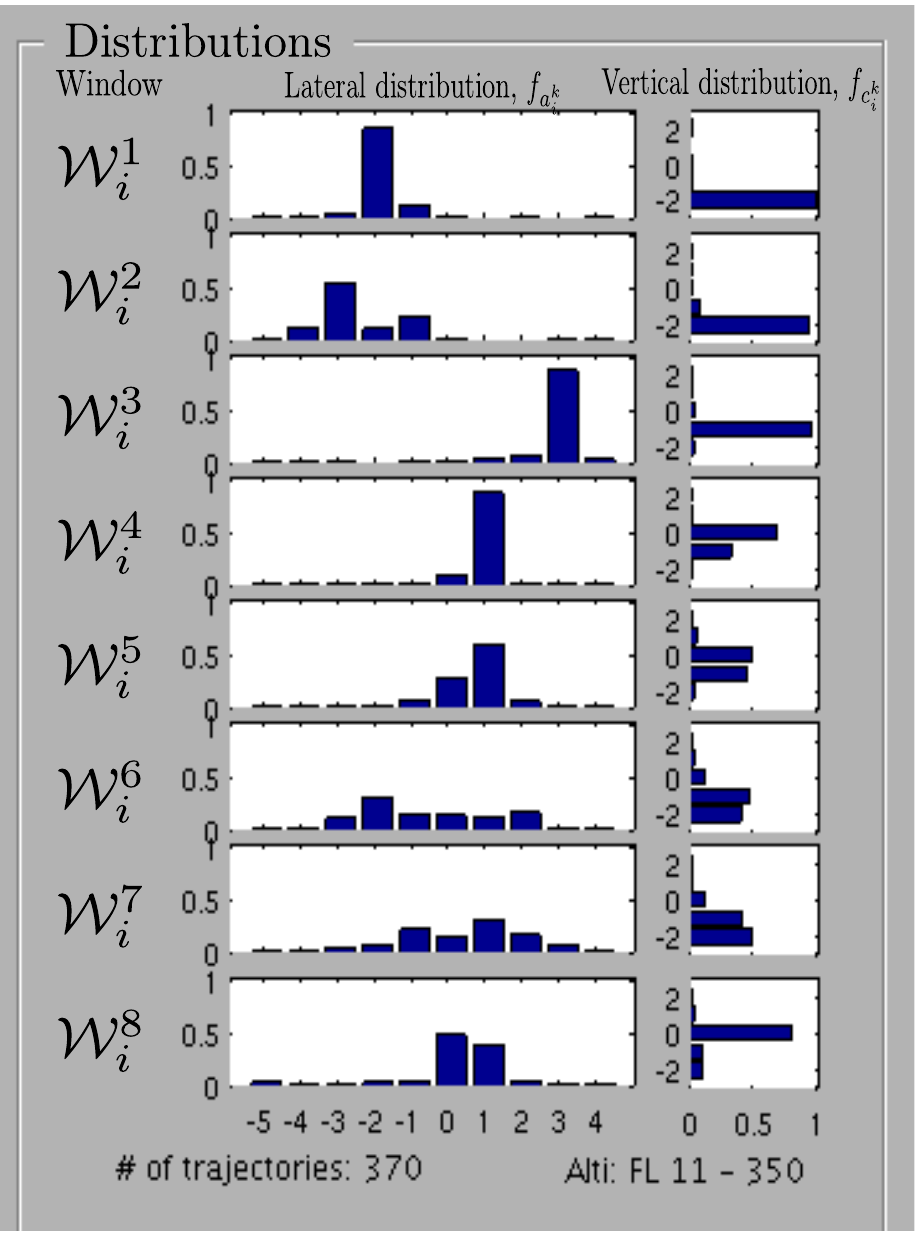}}
\end{minipage}
\caption{Lateral and vertical  distributions of aircraft positions at a window within a flow.}\label{fig:flowsExplorator}
\end{figure}

\begin{figure}[htbp]
 \centering
\includegraphics{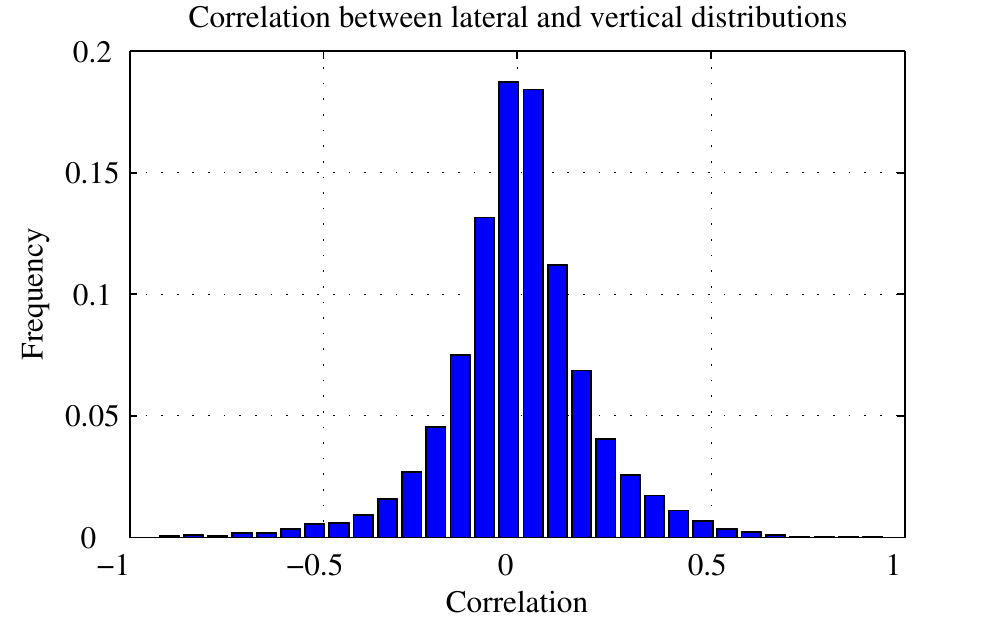}
\caption{Correlation between lateral and vertical distributions of aircraft positions at the windows.}\label{fig:corrXZ}
\end{figure}

Figure~\ref{fig:exampleflow} presents an example cluster, with corresponding trajectories, windows, and tube. The vertical and lateral aircraft distributions for a specific window (resp. for each window $\mathcal{W}_{i}^{k}$) are illustrated lateral Figure~\ref{fig:flow_distribution}. The vertical and horizontal distributions appear to be decorrelated.  For the generative model, the decorrelation of vertical and lateral distribution are assumed.  To support this hypothesis, Figure~\ref{fig:corrXZ} shows a histogram of correlation values between the vertical and lateral distributions, for each of the 8 windows for the 685 flows. The distribution of the correlation values between the vertical and lateral distributions is narrow, and centered in 0.  For 90\% of all cluster windows, the absolute value of the correlation is less than 0.31. This allows the parameterized generative model to use the approximation that the vertical and lateral distribution of aircraft trajectories are independent.

\subsection{Speed distribution}
\begin{figure}[htbp]
\includegraphics{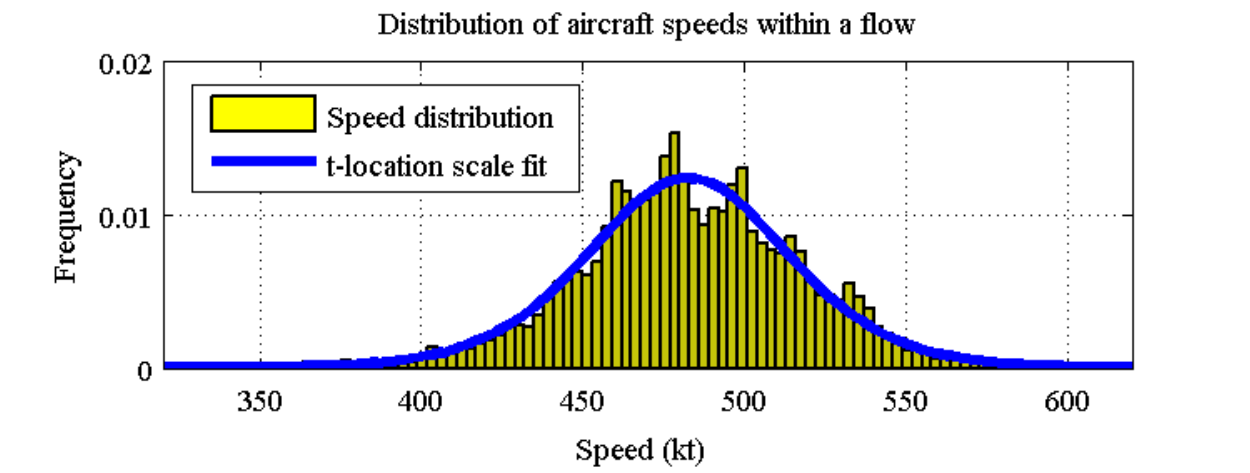}
\caption{Distribution of aircraft speeds for the busiest flow.}\label{fig:speedOneFlow}
\end{figure}

Within a given flow, aircraft fly at different speeds.  Factors for such variation include differences in aircraft and engine type, aircraft weight, weather, wind, and pilot preferences. Analysis of the flows was conducted to determine a distribution to best model  aircraft speeds.  Ultimately, the $t$-location scale probability distribution was selected.  The probability distribution, $f_{V_{i}}$, for the aircraft speeds in flow $\FF_i$ is
\begin{equation}\label{eq:probaDist}
f_{V_{i}}(x) = \frac{\Gamma(\frac{\nu+1}{2})}{\sigma\sqrt{\nu\pi}\Gamma(\frac{\nu}{2})}\left[\frac{\nu+(\frac{x-\mu}{\sigma})^2}{\nu}\right]^{-(\frac{\nu+1}{2})},
\end{equation}
where $\mu$ is the location parameter, $\sigma >0$ is the scale parameter, $\nu >0$ is the shape parameter, and $\Gamma$ is the standard gamma probability distribution. If $x$ has a t-scale distribution with parameters $\mu, \sigma$ and $\nu$, then $\frac{x-\nu}{\sigma}$ has a Student's $t$-distribution with $\nu$ degrees of freedom. Figure~\ref{fig:speedOneFlow} presents an example of an empirical probability density function for aircraft speeds within a major flow, as well as the t-location scale fit for the empirical distribution.

\subsection{Aircraft arrival process}
The clustering method provides a spatial representation of the flows. To obtain a complete traffic model it is necessary that the traffic rate into each flow is appropriately described from historical data. Due to the sparsity of aircraft arrivals into a specific flow, it is difficult to obtain meaningful and verifiable distributions for aircraft arrival times for individual flows. Therefore, the aircraft arrival process is analyzed over the entire airspace region (i.e. the center level)  and proportioned amongst the flows.  Because the number of aircraft entering the airspace region is a function of the time of the day, the objective of this section is to apply the $\chi^2$ test to determine if the aircraft arrival process can be modeled as a non-homogeneous exponential  process.  Additionally, to strengthen the modeling process, similar hypothesis testing is applied to the data to check for a Poisson counting process.  While the Poisson counting process implies an exponential inter-arrival distribution, and vis versa, testing of arrival counting process yields slightly different results.  However, over the vast majority of the day, hypothesis testing reveals that both airspace models are appropriate.

For an arbitrary homogeneous Poisson counting process, $\rv{N}$, defined by parameter $\lambda$ [events/minute], the probability distribution is given by \cite{ross}:
\begin{equation}
P(\rv{N}(t)-\rv{N}(t-s)=k)=\frac{(\lambda)^k}{k!}e^{-\lambda s}.
\label{eq:poisson}
\end{equation}
The associated cumulative distribution for the random variable $\rv{I}$ representing event inter-arrival times, is given by:
\begin{equation}
P(\rv{I}<a)=1-e^{-\lambda a}.
\label{eq:exponential}
\end{equation}

To drive the modeling process, each day is divided into 15 minute increments, during which time, the aircraft arrival process is assumed to be static and follow exponential inter-arrival times or a Poisson process.  The $\chi^2$ goodness-of-fit test is applied to all 15 minute time intervals for both the exponential inter-arrival time distributions, as well as the Poisson counting process.  The null hypothesis declares that the inter-arrival times into the sector are exponential, with rate $\lambda^{j}$; or in the case of the counting process, that the distribution follows a Poisson process with parameter $\lambda^{j}\tau$.  The parameter, $\lambda^{j}\tau$, is the average number of aircraft arrivals during the $j^{th}$ time interval of length $\tau$, over all similar days of the week. In this case, $\tau = 15 min$. For a $\chi^2$-test result of 1, the null hypothesis is rejected, implying that it is unlikely that the inter-arrival times  follows an exponential distribution (or Poisson counting process) with rate $\lambda^j$ ($\lambda^{j}\tau$); otherwise, the hypothesis cannot be rejected. The $\alpha$ significance level was set to 0.05, which is a standard value for this test.  For the exponential distribution, the bin-size for  $\chi^2$-test is selected to be 2 seconds.

\begin{figure}[htbp]
\centering
\subfigure[Evolution of the rate parameter (Zulu).]{\includegraphics{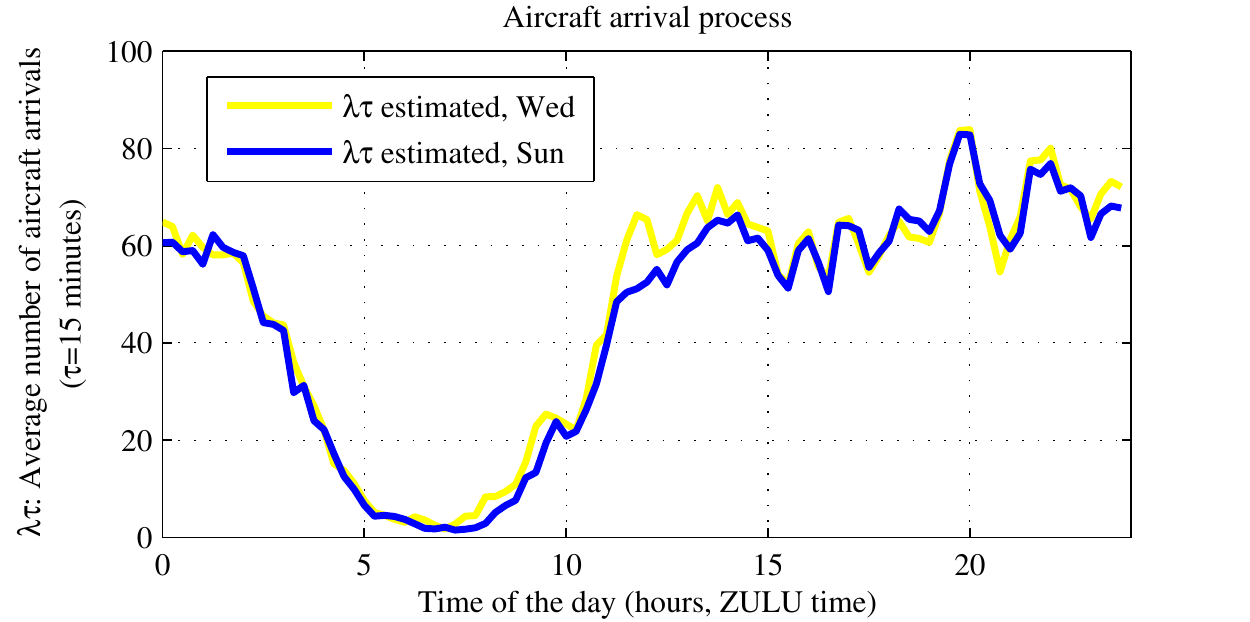}\label{fig:sectorPoissonProcess_2}}
\subfigure[$\chi^2$ test results for Wednesdays.]{\includegraphics{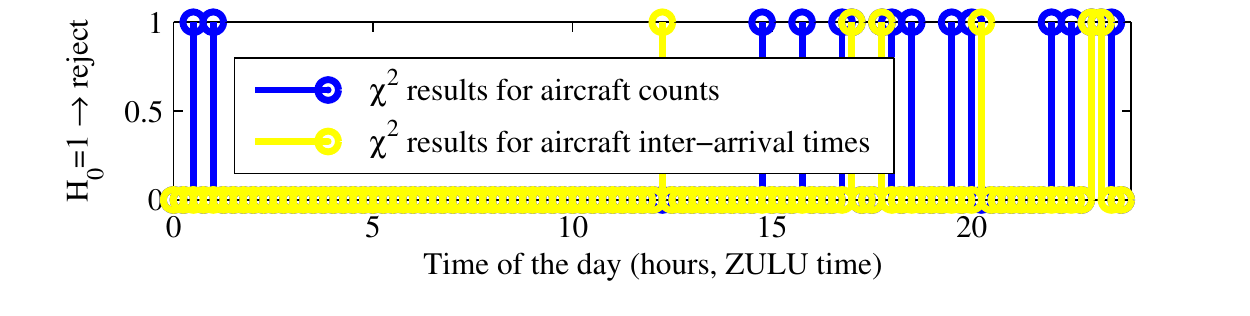}\label{fig:chi2_wed}}
\subfigure[Example emprical and modeled expoential distributions.]{\includegraphics{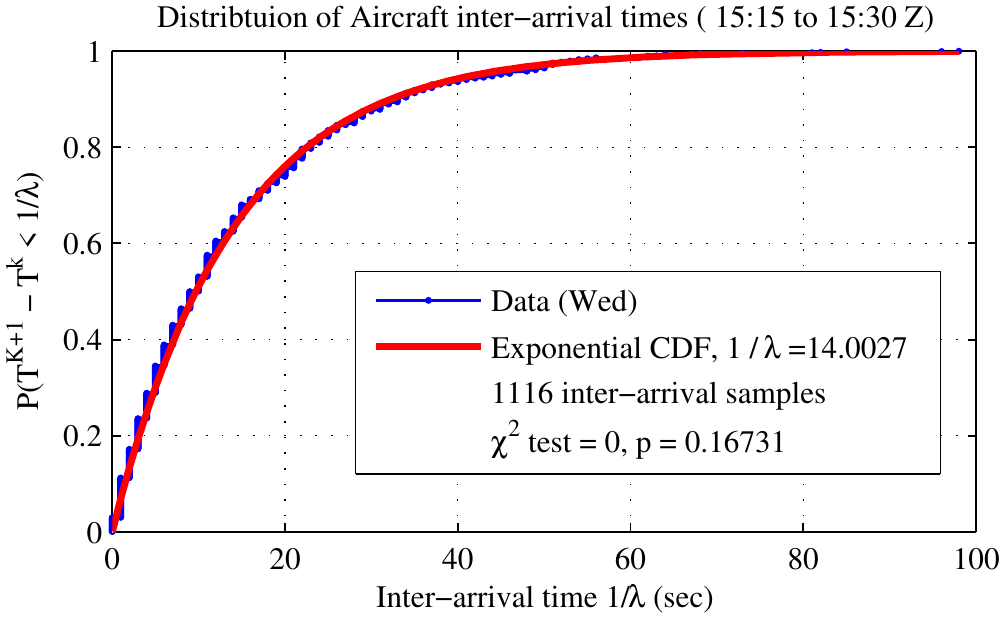}\label{fig:sectorPoissonProcess_1}}
\caption{Aircraft arrival process into the center.}\label{fig:sectorPoissonProcess}
\end{figure}

Figure~\ref{fig:sectorPoissonProcess_2} illustrates the traffic variation over the day according to the best-fit 15 minute parameter $\lambda^j \tau$ for the days Wednesday and Sunday.  Over the vast majority of the 24 hour time span considered, for Wednesdays (although the same results hold for all other days), the null hypothesis is not rejected for either the exponential distribution nor the Poisson process, as illustrated in Figure~\ref{fig:chi2_wed}.   That is, the arrival process for each 15 minute interval follows a best-fit exponential distribution.  The differences between the values of $\lambda^{j}\tau$ for Wednesdays and Sundays indicate daily changes in air traffic demand.

As an example, for the time period between 15:15 and 15:30 Zulu time, over all Wednesday of the week, comparison the distribution of aircraft inter-arrival times into the sector, with the best-fit exponential distribution, shows a close fit, as illustrated in Fig.~\ref{fig:sectorPoissonProcess_1}.

The aircraft inter-arrival process into the center is therefore modeled as a non-homogeneous poisson process over the complete day, with static rate parameters for any 15 minute interval. The rate parameter changes with the time of the day, and day of the week. Additionally, it is likely that seasonal changes exist.  However for the 4 months considered, any long-term variations over same days of the week are small.  While, the $\chi^2$ test does not validate that aircraft arrive with an exponential distribution, the test does verify a specific property, enabling modeling of aircraft inter-arrival times into the airspace as exponential.  Verifying that the air traffic arrival process into an en route airspace can be modeled as an exponential distribution supports an assumption found extensively throughout the literature (e.g. see~\cite{clarke:2009,moreau2005,Salaun_DASC:2009}).

Similar modeling statements concerning the arrival process into each flow is not possible, as the number of aircraft arrivals per flow per 15 minute period is low, and thus fails to yield strong results for $\chi^2$ testing. For modeling purposes, we assume that the flows are independent, and so are their arrival processes.  The arrival process for the greater airspace is proportioned to according to traffic volume along each flow. The rate parameter for the flow $\mathcal{F}_i$ during time period $j$ is $\lambda^j_{i} = \pi_{i}^j \lambda^j$, where $\pi_{i}^j$ is given by
\begin{equation}
\pi_{i}^j = \frac{\displaystyle\sum_{d=1}^{30}x^j_{d,i}}{\displaystyle\sum_{d=1}^{30}x^j_{d}},
\end{equation}
with $x^j_{d,i}$ being the count of aircraft entering the center during the time period $j$, on data day $d$, and identified flying on flow $\mathcal{F}_i$.

%

\subsection{Outlier model}
\begin{figure}[htbp]
\includegraphics[width=0.9\textwidth]{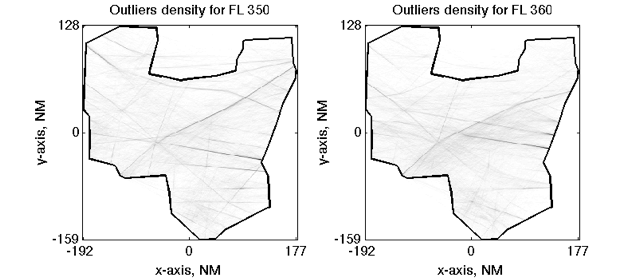}
\caption{Density of outliers at FL 350 and FL 360.}\label{fig:outliersXray}
\end{figure}

Outliers are aircraft that are not clustered into flows, and they represent approximately 20\% of the traffic. To model outlier trajectories, density maps of aircraft positions are generated.  The airspace is discretized into 1 NM squares and 1000 ft high cubes.  By spatially interpolating all aircraft trajectories, the total number of flights passing through a particular cube is determined.  To generate a probability distribution, the aircraft counts in each cube are normalized between 0 and 1 over the entire airspace.  Figure~\ref{fig:outliersXray} presents density plots for the location of outliers at FL 350 and FL 360.  In the figure, black indicates a high density of aircraft, and white a low density. With exception to the location where major flows pass, the distribution of outliers over the airspace at a given flight level is relatively uniform.  The spatially probability distribution of outliers defines an outlier probability model, $f_{O}(P)$.

Drawing an analogy with the Kalman filter, the clusters can be interpreted to be the main signal, and the outliers as a noise injected into the signal.

\subsection{Flow and outlier reconstruction}
Over the last two sections, clustering and outlier analysis provided models for the spatial distribution of aircraft through airspace.  Furthermore, the vertical and lateral distribution of aircraft positions at each window provides greater granularity in the spatial distribution of aircraft.  Coupled with a model for aircraft speeds, and aircraft inter-arrival times into the airspace, an air traffic model is generated.  The benefit of such a model is that it can be parameterized to test `what-if' scenarios without reconstruction of the traffic model.  That is, adjustments to the model properties: flow rates, mean aircraft speeds, trajectory windows, or flow trajectories can be made quickly, thereby defining a parameterized generative model.  The information provided by parameterized generative model is then used to construct the proximity maps proposed in this paper.

 \section{Generation of aircraft proximity maps}
\label{sec:proximity}

Modeling airspace via aircraft flows and outliers enables the generation of the presence, conflict, and outlier proximity maps, whose computation is now presented. These maps use the information generated by the parameterized generative model detailed in the previous sections. We focus first on the presence and conflict maps, which consider only aircraft from modeled flows. Then, we will introduce outlier proximity maps.

\subsection{Definition}
 Consider the East-North-Up frame $\mathcal{R}_0=(\overrightarrow{a_0},\overrightarrow{b_0},\overrightarrow{c_0})$ defined in Section~\ref{sec:data} and fixed with respect to the airspace $\mathcal{A}$; and consider any point $P=(x_p,y_p,y_p)$ of the airspace. Given a model of air traffic parameterized, the objective is to compute maps based on the probability $\mathbf{P}_1(P)$ that at least one aircraft flies in the proximity of a point $P$ and the probability $\mathbf{P}_2(P)$ that at least two aircraft from different flows fly in the proximity of a point $P$. These probabilities will help to generate the presence maps (related to $\mathbf{P}_1(P)$) and the conflict map (related to $\mathbf{P}_2(P)$).

\begin{figure}[htbp]
 \centerline{
 \qquad\qquad\qquad \includegraphics[width=9cm]{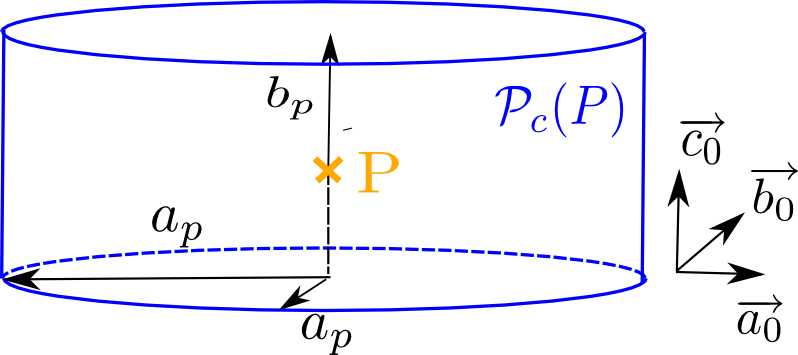}}
 \caption{Proximity volume $\mathcal{P}_c(P)$ for a given point $P$ of the airspace.}\label{def_proximity_area}
\end{figure}

  The probability $\mathbf{P}_1(P)$ can be interpreted as the probability of presence of aircraft, and $\mathbf{P}_2(P)$ can be related to probability of conflict. The shape of the proximity volume is arbitrary but we consider first the proximity volume as the full cylinder $\mathcal{P}_c(P)$ centered on $P$, and defined as follows:
\begin{equation}\label{eq:def_proximity_area}
\mathcal{P}_c(P)=\{(x,y,z)\in \mathcal{A}\ |\ (x-x_p)^2+(y-y_p)^2\leq a_p^2\quad \text{and}\quad (z-z_p)^2\leq b_p^2\},
\end{equation}
with $a_p,b_p>0$, as illustrated in Fig.~\ref{def_proximity_area}.  Choosing $a_p=2.5$ NM, ${b_p=1000 \text{ft} = 0.165 \text{NM}}$ reflects the notion of conflict between aircraft: if two aircraft are simultaneously located inside the cylinder, they are in conflict. 
For any point $P$, the probability $\mathbf{P}_1(P)$ (resp. $\mathbf{P}_2(P)$) is then the probability that at least one (resp. two) aircraft from different flows are located inside $\mathcal{P}_c(P)$, at any given time. However, considering the proximity volume $\mathcal{P}_c(P)$ given in Eq.~\eqref{eq:def_proximity_area} is computationally too expensive to generate proximity maps in real-time, since the flows are not only a single track but a 3D distribution. The cylinder is approximated by its circumscribing rectangular box. The resulting values of the probabilities $\mathbf{P}_1(P)$ and $\mathbf{P}_2(P)$  are then upper bounds on the corresponding values if $\mathcal{P}_c(P)$ had been considered instead.

For a given airspace geometry and flow configuration, two three-dimensional proximity maps, $\mathcal{M}_1$ and $\mathcal{M}_2$, are generated according to the values $\mathbf{P}_1(P)$ and $\mathbf{P}_2(P)$ associated with all points $P$ in the airspace. While the map $\mathcal{M}_1$ is based on the global presence of aircraft from a flow in the airspace, the map $\mathcal{M}_2$ highlights the regions of possible conflict between aircraft from different flows, which are the most sensitive regions for air traffic controllers.

\subsection{Inter-aircraft distance and ``residual distance'' of an aircraft}

 From Section~\ref{sec:data}, a flow $\mathcal{F}_i$ is characterized by, among other parameters, four probability density functions : $f_{a_i^k}$ and $f_{c_i^k}$ for the lateral and vertical distributions of aircraft at the window $\mathcal{W}_i^k$; $f_{\Delta T_i}$ for the inter-arrival time distribution between two consecutive aircraft (for clarity, the index of the day $j$ is omitted in this section); $f_{V_i}$ for the aircraft speed distribution. To create spatial proximity maps, the inter-aircraft distance, i.e. the distance between two consecutive aircraft flying within the same flow $\mathcal{F}_i$, must be introduced, instead of inter-arrival time. It will result a complete knowledge of spatial distribution of aircraft within the flow $\mathcal{F}_i$, along the vectors $(\overrightarrow{a_i^k},\overrightarrow{b_i^k},\overrightarrow{c_i^k})$.

\begin{figure}[htbp]
  \centering \subfigure[Inter-arrival distance]{\includegraphics[width=7cm]{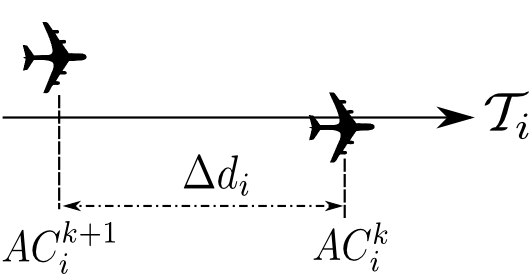}\label{inter_arrival}}\qquad
  \subfigure[Residual distance]{\includegraphics[width=7cm]{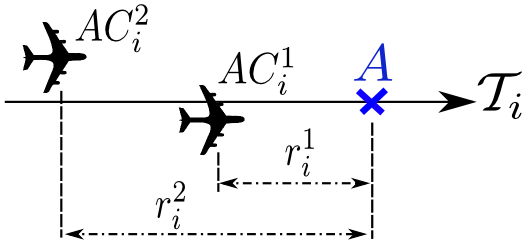}\label{residual distance}}
 \caption{Definitions of the inter-aircraft distance and residual distance.}
 \end{figure}

Denote $\{AC_i^k, k\in\Nset^*\}$ the sequence of aircraft flying in flow $\mathcal{F}_i$.  The longitudinal distance $\Delta d_i$ between two consecutive aircraft of $\mathcal{F}_i$ is defined as the distance between aircraft projection on the track (or centroid) $\mathcal{T}_i$ (see definition in Section~\ref{sec:data} and depicted in Fig.~\ref{inter_arrival}) of two consecutive aircraft  $(AC_i^k, AC_i^{k+1})$. The inter-aircraft distance is a random variable such that $\Delta d_i=v_i\Delta t_i$, and has associated probability density function $f_{\Delta D_i}$. Assuming that the two random variables $v_i$ and $\Delta t_i$ are independent, $f_{\Delta D_i}$ can be determined from the knowledge of $f_{V_i}$ and $f_{\Delta T_i}$. Let $\Delta d_i^{m}$ be the mean inter-aircraft distance.

The notion of ``residual distance'' of the aircraft is now introduced, as it is a key to the construction of the proximity maps. The notion of ``residual distance'' for aircraft is similar to the concept of ``residual life'' (or ``excess life'') in renewal theory (see~\cite{ross} for further details), and it is defined as follows: consider any point $A$ located on the track $\mathcal{T}_i$ of flow  $\mathcal{F}_i$ (see Fig.~\ref{residual distance}). Observing traffic at time $t_0 = 0$ (we assume that the process starts at time $t=-\infty$), let $AC_i^k$ be the \emph{next} $k^\text{th}$ aircraft to cross $A$ at any time $t\geq t_0$. The ``residual distance''~$r_i^k$ of the aircraft $AC_i^k$ is the distance between the projection of the current position of aircraft $AC_i^k$ onto $\mathcal{T}_i$ and $A$. Especially, $r_i^1$ is the distance between the current position of aircraft $AC_i^1$ and $A$. The variable $r_i^1$ is called the ``first'' residual distance because it is related to the first aircraft to cross $A$. From~\cite{ross}, the PDF of the residual distance~$r_i^k$ of the aircraft $AC_i^k$, ${f_{R_i^k}(r_i^k)}$, can be derived from the PDF of the inter-aircraft distance, $f_{\Delta D_i}(\Delta d_i)$, and is given by the following expression:
\begin{align}\label{eq:pdf_r_1}
f_{R_i^1}(r_i^1)&=
\frac{1}{\Delta d_i^{m}}\bigl(1-F_{\Delta D_i}(r_i^1)\bigr),\\
f_{R_i^k}&=f_{R_i^{k-1}}*f_{\Delta D_i},\quad k\geq2,\label{eq:pdf_r_k}
\end{align}
where $F_{\Delta D_i}$ is the cumulative distribution function of the inter-aircraft distance associated to $f_{\Delta D_i}$, and $*$ represents the convolution operator.

To illustrate the notion of ``residual distance'', consider a straight flow $\mathcal{F}_i$ and assume the same constant velocity for all aircraft within the flow, given by the maximum velocity estimate of the speed distribution. From Section~\ref{sec:data}, it is assumed that the inter-arrival time distribution $f_{\Delta T_i}$ is an exponential distribution, characterized by the arrival rate $\lambda_i$. The PDF $f_{\Delta D_i}$ is then also an exponential distribution, characterized by the mean inter-aircraft distance $\Delta d_i^m$:
 \begin{equation}\label{eq:expon_distrib}
 f_{\Delta D_i}(\Delta d_i)=\left\{
\begin{array}{ll}
 \frac{1}{\Delta d_i^m}e^{-\frac{1}{\Delta d_i^m}\Delta d_i}\quad & \text{when}\quad \Delta d_i\geq 0\\
0\quad & \text{when}\quad \Delta d_i<0.
\end{array}\right.
\end{equation}
From Eq.~\eqref{eq:pdf_r_1}, the probability density function of the ``first'' residual distance, ${f_{R_i^1}}$, can then be written as:
 \begin{equation}\label{eq:expon_distrib_r_1}
 f_{R_i^1}(r_i^1)=\left\{
\begin{array}{ll}
 \frac{1}{\Delta d_i^m}e^{-\frac{1}{\Delta d_i^m}r_i^1}\quad & \text{when}\quad r_i^1\geq 0\\
0\quad & \text{when}\quad r_i^1<0,
\end{array}\right.
\end{equation}
which is similar to $f_{\Delta D_i}$ for the considered example. For any $k^{th}$ aircraft $AC_i^k$, the probability density function of its residual distance $r_i^k$, ${f_{R_i^k}}$, can be then derived from Eqs.~\eqref{eq:pdf_r_k} and~\eqref{eq:expon_distrib_r_1}.

\subsection{Intersection volume}

\begin{figure}[htbp]
 \centerline{
 \includegraphics[width=11cm]{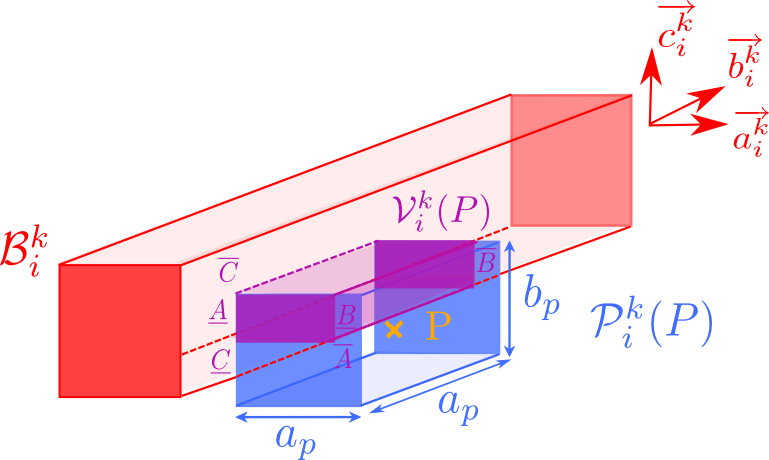}}
 \caption{Definition of the intersection volume $\mathcal{V}_i^k(P)$.}\label{def_v_i_k}
\end{figure}

As mentioned above, considering a cylinder as the proximity volume requires too many computations to generate results for a realistic region of the airspace in real-time (i.e. in a few seconds). 
Therefore, the proximity cylinder $\mathcal{P}_c(P)$ is approximated by a circumscribing box, whose geometry depends of the box $\mathcal{B}_i^k$ considered. Indeed, given any $k^{th}$ box $\mathcal{B}_i^k$ of flow $\mathcal{F}_i$ and any point $P\in\mathcal{A}$, the corresponding proximity box $\mathcal{P}_i^k(P)$ is defined such that (see Fig.~\ref{def_v_i_k}):
\begin{itemize}
\item its center is $P$,
\item its dimensions are $a_p\times a_p \times b_p$ (5NM$\times$5NM$\times$0.165NM),
\item it is aligned with the centroid of $\mathcal{B}_i^k$.
\end{itemize}

To compute the probabilities $\mathbf{P}_1(P)$ and $\mathbf{P}_2(P)$, the first step is to determine the intersection volume $\mathcal{V}_i^k(P)$, defined to be the intersection of any box $\mathcal{B}_i^k$ with $\mathcal{P}_i^k(P)$, i.e. $\mathcal{V}_i^k(P)=\mathcal{B}_i^k \bigcap \mathcal{P}_i^k(P)$ as depicted in Fig.~\ref{def_v_i_k}. Considering both the dimensions of the proximity box (5NM$\times$5NM$\times$0.165NM), and the fact that aircraft trajectories are slanted just a few degrees (due to the physical constraints of the aircraft flight dynamics), it is approximated for computational simplicity that the considered box $\mathcal{B}_i^k$ is horizontal around $\mathcal{P}_i^k(P)$.

Consider the frame $\mathcal{R}_i^k=(\overrightarrow{a_i^k},\overrightarrow{b_i^k},\overrightarrow{c_i^k})$ corresponding to the box $\mathcal{B}_i^k$, with $P$ as the geometric center. Since $\mathcal{R}_i^k$ is aligned with the box $\mathcal{B}_i^k$, it is also aligned with $\mathcal{V}_i^k(P)$. The bound of the intersection volume $\mathcal{V}_i^k(P)$ is then defined by its entry and exit coordinates along the three axes: $\underline{A}$ and $\overline{A}$ along $\overrightarrow{a_i^k}$, $\underline{B}$ and $\overline{B}$ along $\overrightarrow{b_i^k}$, and $\underline{C}$ and $\overline{C}$ along $\overrightarrow{c_i^k}$. These values can be easily determined from the equations of the proximity volume~$\mathcal{P}_i^k(P)$ and of the box~$\mathcal{B}_i^k$.

\subsection{Presence maps}
\label{sec:first_order_map}
Presence maps consider the probability that one or more aircraft is located near a point $P$, i.e. inside the considered proximity volume.  Given any point $P\in\mathcal{A}$, consider first a single box $\mathcal{B}_i^k$, and the proximity box $\mathcal{P}_i^k(P)$ . Let $\mathbf{P}_1(P,\mathcal{B}_i^k)$ be the probability that at least one aircraft $AC_i$ from box $\mathcal{B}_i^k$ be inside $\mathcal{P}_i^k(P)$. Since the three probability distributions for the presence of an aircraft along each axis are assumed to be independent (see Section~\ref{sec:data}), the probability $\mathbf{P}_1(P,\mathcal{B}_i^k)$ can be written
\begin{equation}\label{eq:def_proba_1}
\mathbf{P}_1(P,\mathcal{B}_i^k)=\mathbf{P}(\exists\ AC_i \in \mathcal{P}_i^k(P))=\mathbf{P}(\exists\ AC_i \in [\underline{A} \overline{A}]).\mathbf{P}(\exists\ AC_i \in [\underline{B} \overline{B}]).\mathbf{P}(\exists\ AC_i \in [\underline{C} \overline{C}]).
\end{equation}
So far, only the lateral and vertical PDFs at the windows $\mathcal{W}_i^k$ (i.e. at the entrance of the boxes $\mathcal{B}_i^k$) have been defined and determined in Section~\ref{sec:data}: $f_{a_i^k}$ and $f_{c_i^k}$. Since the intersection volume $\mathcal{V}_i^k(P)$ is not necessarily near $\mathcal{W}_i^k$, the aircraft spatial distributions within the flow $\mathcal{F}_i$ must be extended around the points $\underline{A}$,$\underline{C}$. Therefore, a linear interpolation of the lateral (resp. vertical) distributions $f_{a_i^k}$ and $f_{a_i^{k+1}}$ (resp. $f_{c_i^k}$ and $f_{c_i^{k+1}}$) between the windows $\mathcal{W}_i^k$ and $\mathcal{W}_i^{k+1}$ is used to get $f_{a_i^k/\underline{A}}$ (resp. $f_{c_i^k/\underline{C}}$), i.e. the lateral (resp. vertical) aircraft distribution at the point $\underline{A}$ (resp. $\underline{C}$). The linear interpolation process approximates the ``smooth'' variation of aircraft presence density within the flow, preserving the nature of a probability density function.

Any term in the right-hand side of Eq.~\eqref{eq:def_proba_1} can be computed as a function of the probability density functions previously defined, $f_{a_i^k/\underline{A}}$, $f_{c_i^k/\underline{C}}$, and $f_{R_i^1}$ :
\begin{align}
\mathbf{P}(\exists\ AC_i \in [\underline{A} \overline{A}])&=\int_{\underline{A}}^{\overline{A}}f_{a_i^k/\underline{A}}(x)\,dx,\label{eq:def_proba_a_i_k}\\
\mathbf{P}(\exists\ AC_i \in [\underline{C} \overline{C}])&=\int_{\underline{C}}^{\overline{C}}f_{c_i^k/\underline{C}}(z)\,dz,\label{eq:def_proba_b_i_k}\\
\mathbf{P}(\exists\ AC_i \in [\underline{B} \overline{B}])&=\int_{0}^{\underline{B}\overline{B}}f_{R_i^1}(y)\,dy,\label{eq:def_proba_c_i_k}
\end{align}
where $\underline{B}\overline{B}=\norm{\overrightarrow{\underline{B}\overline{B}}}$. The probability~$\mathbf{P}_1(P,\mathcal{B}_i^k)$ can then be determined from Eqs.~\eqref{eq:def_proba_1}--\eqref{eq:def_proba_c_i_k}.

The previous definitions and equations can naturally be extended to the $l_1$ boxes $\mathcal{B}_i^k$ of the flow $\mathcal{F}_i$. It leads to introduce the probability, $\mathbf{P}_1(P,\mathcal{F}_i)$, that there is at least one aircraft $AC_i$ from flow $\mathcal{F}_i$ flying at proximity of the point $P$. The probability is computed from Eqs.~\eqref{eq:def_proba_1}--\eqref{eq:def_proba_c_i_k} taking the $l_1$ boxes $\mathcal{B}_i^k$ into account.

Consider now the entire airspace $\mathcal{A}$ modeled with $N$ independent flows. The presence probability of at least one aircraft from a flow about any point $P$, $\mathbf{P}_1(P)$, can then be written
\begin{equation}\label{eq:def_proba_first_order}
\mathbf{P}_1(P)=1-\mathbf{P}(\forall i\in[1,N],\ \nexists\ AC_i \in \mathcal{P}_i(P))=1-\prod_{i=1}^N \mathbf{P}(\nexists\ AC_i \in \mathcal{P}_i(P))=1-\prod_{i=1}^N \bigl( 1- \mathbf{P}_1(P,\mathcal{F}_i)\bigr).
\end{equation}

\subsection{Conflict maps}
\label{sec:second_order_map}

Conflict maps determine the probability that at least two aircraft from different flows are simultaneously in the same region of the airspace. More formally, the probability $\mathbf{P}_2(P)$, that at least two aircraft from different flows are in the proximity volume surrounding $P$, is written :
\begin{equation*}\label{eq:second_n_ac}
\mathbf{P}_2(P)=\mathbf{P}\bigl(\exists (i,j) \in  [1,N]^2,i\neq j,\ \text{s.t.}\ (AC_i \in \mathcal{P}_i(P))\ \&\ (AC_j \in \mathcal{P}_j(P)\bigr).
\end{equation*}
Using the probability $\mathbf{P}_1(P)$ defined in Section~\ref{sec:first_order_map}, $\mathbf{P}_2(P)$ can be written
\begin{equation}
\begin{aligned}
\mathbf{P}_2(P)&=\mathbf{P}_1(P)-\mathbf{P}(\exists i\in [1,N],\ (AC_i \in \mathcal{P}_i(P))\ \&\ (AC_j \notin \mathcal{P}_j(P),\forall j\neq i))\nonumber\\
&=\mathbf{P}_1(P)-\sum_{i=1}^N\mathbf{P}_{1}(P,\mathcal{F}_i)\prod_{j=1,j\neq i}^N(1-\mathbf{P}_1(P,\mathcal{F}_j)).\label{eq:second_n_ac_dev}
\end{aligned}
\end{equation}

\subsection{Outlier proximity maps}
\label{sec:outliers_map}

Presence and conflict maps consider only aircraft that belong to flows. For a complete modeling of the traffic, outliers -- non clustered aircraft -- should be considered because they account for 20\% of the overall traffic and may have an important influence on the dominant traffic. The spatial distribution of the outliers, $f_O(P)$, is given by the parameterized generative model presented in Section~\ref{sec:gen_model}. This leads to the introduction of outlier proximity maps, denoted $\mathcal{M}_O$. Outliers proximity maps estimate the influence of outliers on the modeled flows that represent the majority of traffic inside the airspace under consideration. Therefore the probability $\mathbf{P}_O(P)$, that one aircraft from the flows and one outlier are simultaneously in the same neighborhood of any point $P$ of the airspace, is determined. Only one outlier is considered, since the probability of two outliers to be in proximity too each other is negligible. Using a methodology similar to the one used in Sections~\ref{sec:first_order_map} and~\ref{sec:second_order_map}, $\mathbf{P}_O(P)$ is approximated by
\begin{equation*}
\mathbf{P}_O(P)=\mathbf{P}_1(P).\frac{1}{a_P^2b_P}\iiint_{\mathcal{P}_O(P)}f_O(P')\,d\tau,
\end{equation*}
where $\mathcal{P}_O(P)$ is a parallelogram whose dimensions are $a_P\times a_P\times2b_P$, and whose center is $P$. The spatial distribution of the outliers, $f_O(P)$, can be designed (i.e. simulated) to evaluate the influence of the presence of rogue aircraft (considered as external disturbances) on a ``nominal'' air traffic consisting of the flows in a given region of the airspace: this approach extends the works in~\cite{lee:jgcd_09,prandiniCSP:2009} from individual aircraft to entire aircraft flows.  
 \section{Generation of proximity maps for Cleveland center ZOB}
\label{sec:example}
\begin{figure}[htbp]
\centering
\subfigure[3D view]{\includegraphics[width=0.49\textwidth]{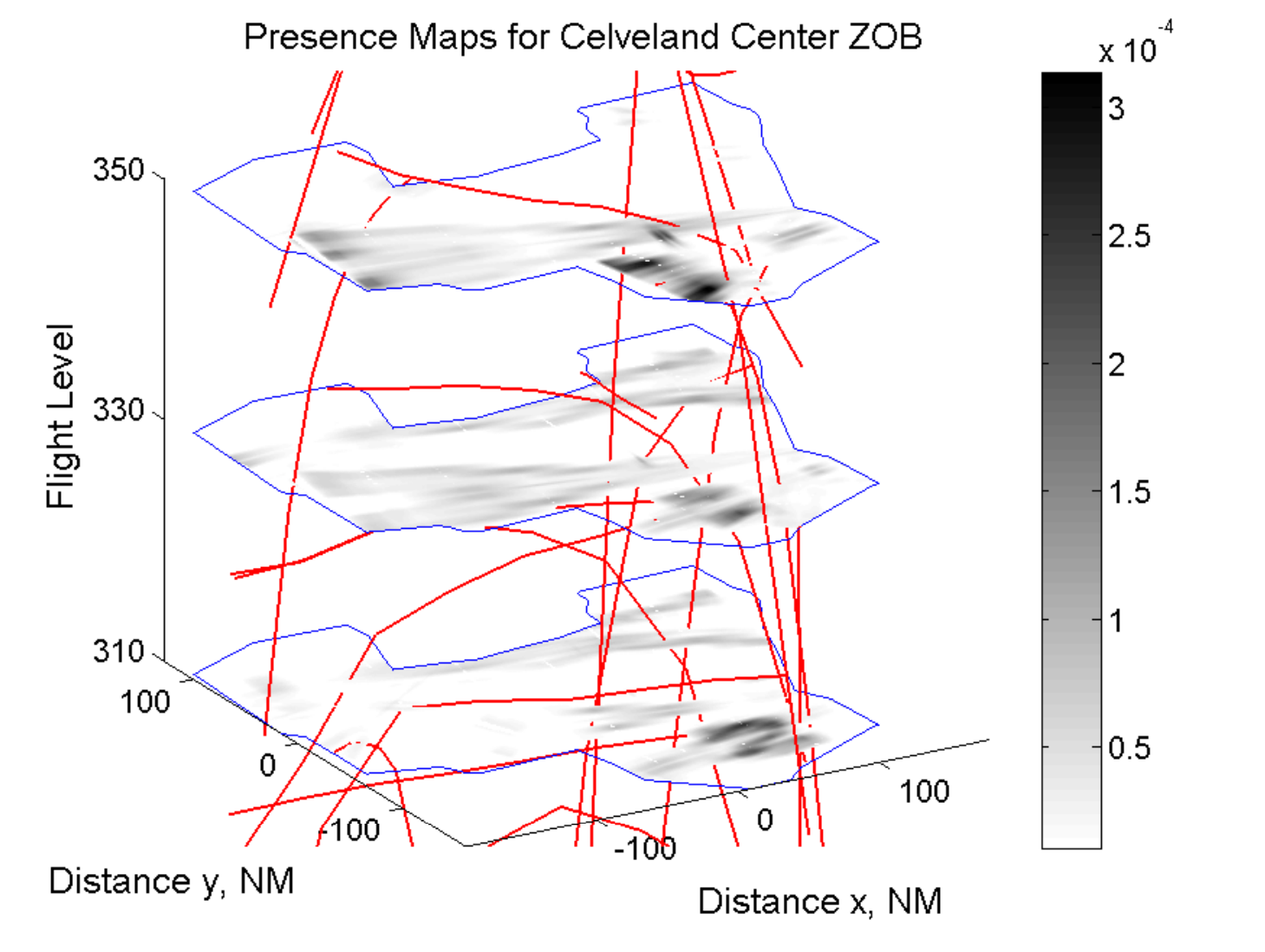}\label{first_map}}
\subfigure[2D view at FL 310]{\includegraphics[width=0.49\textwidth]{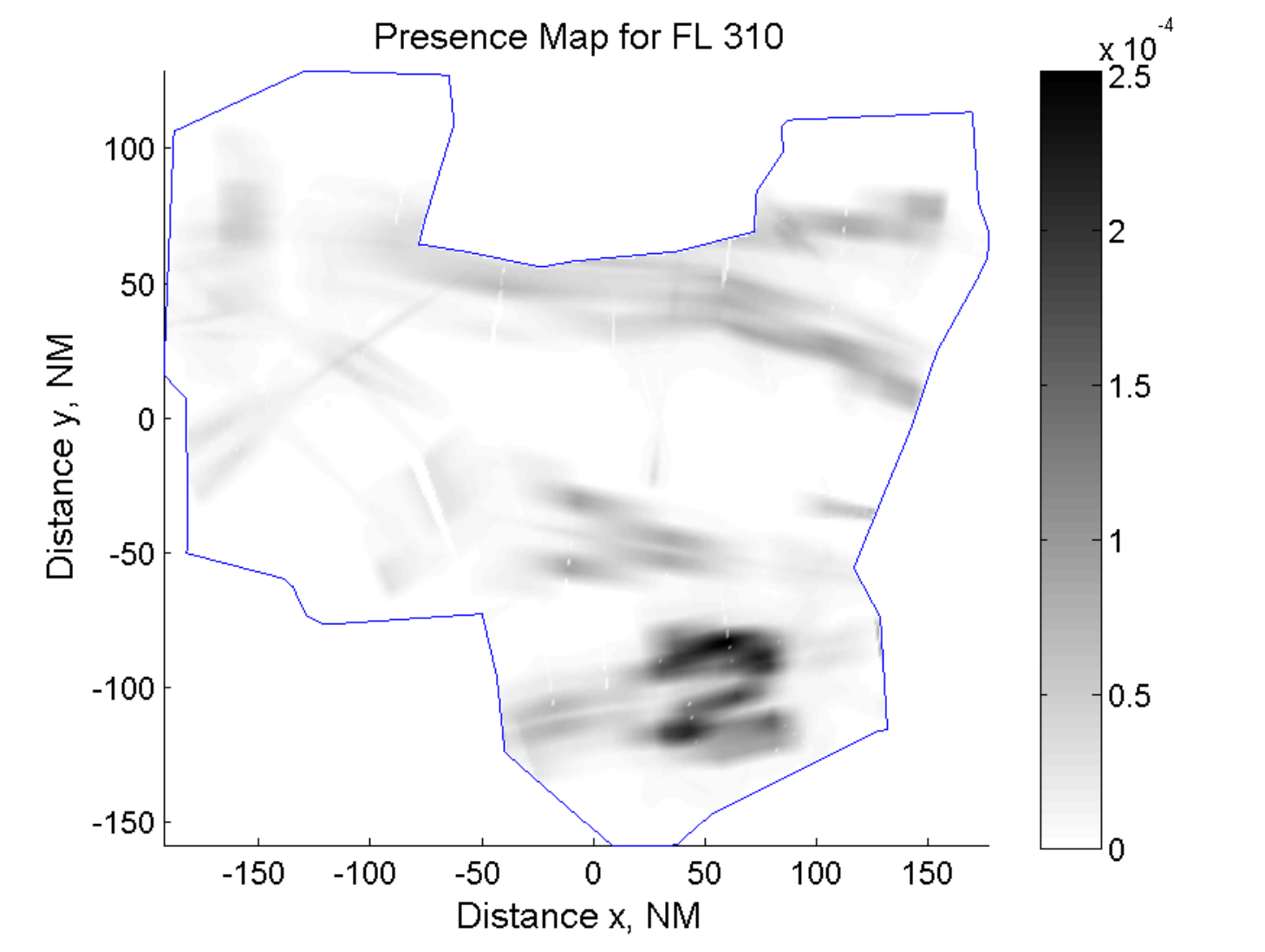}\label{first_map_1}}\\
\subfigure[2D view at FL 350]{\includegraphics[width=0.49\textwidth]{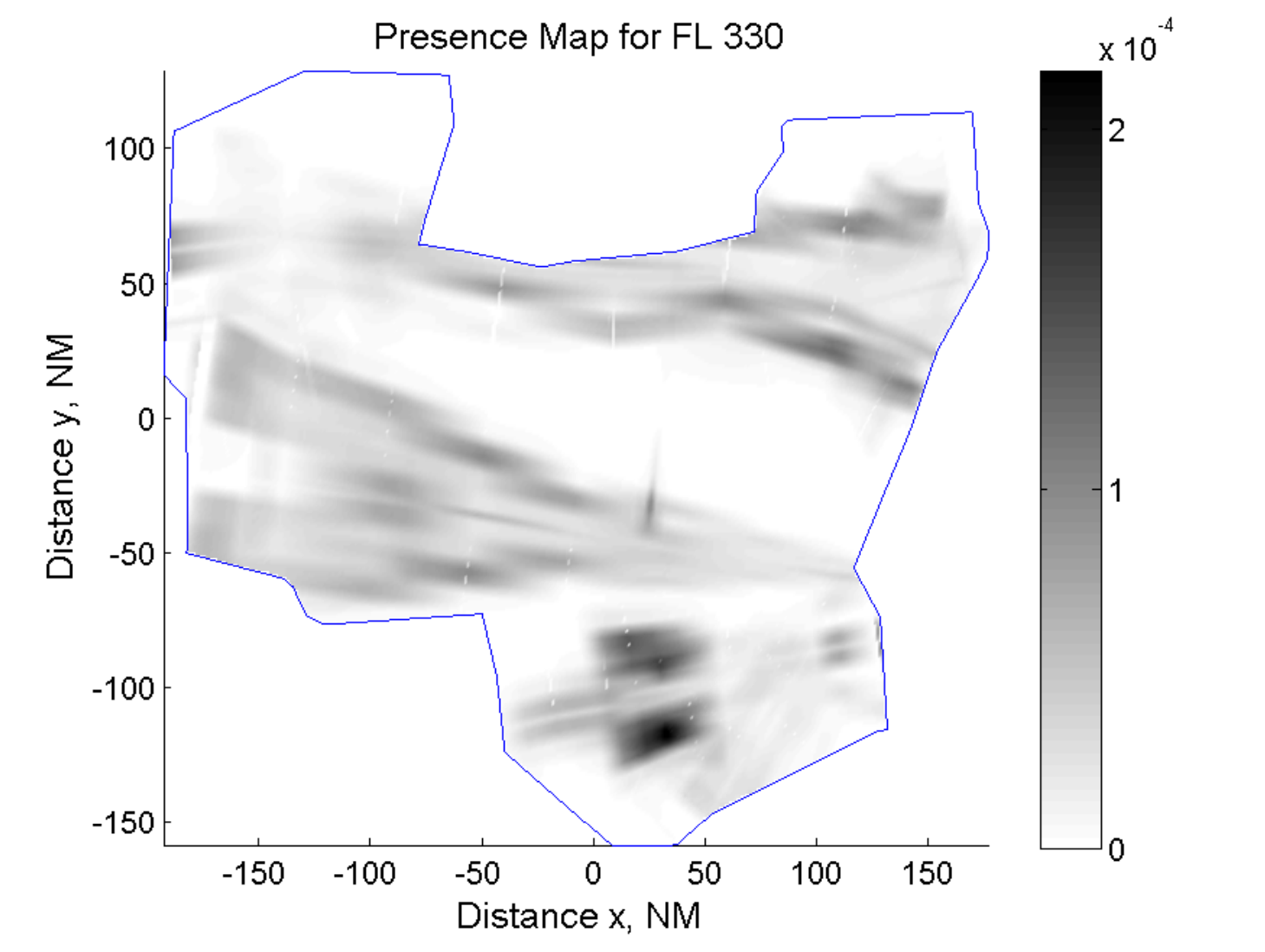}\label{first_map_2}}
\subfigure[2D view at FL 380]{\includegraphics[width=0.49\textwidth]{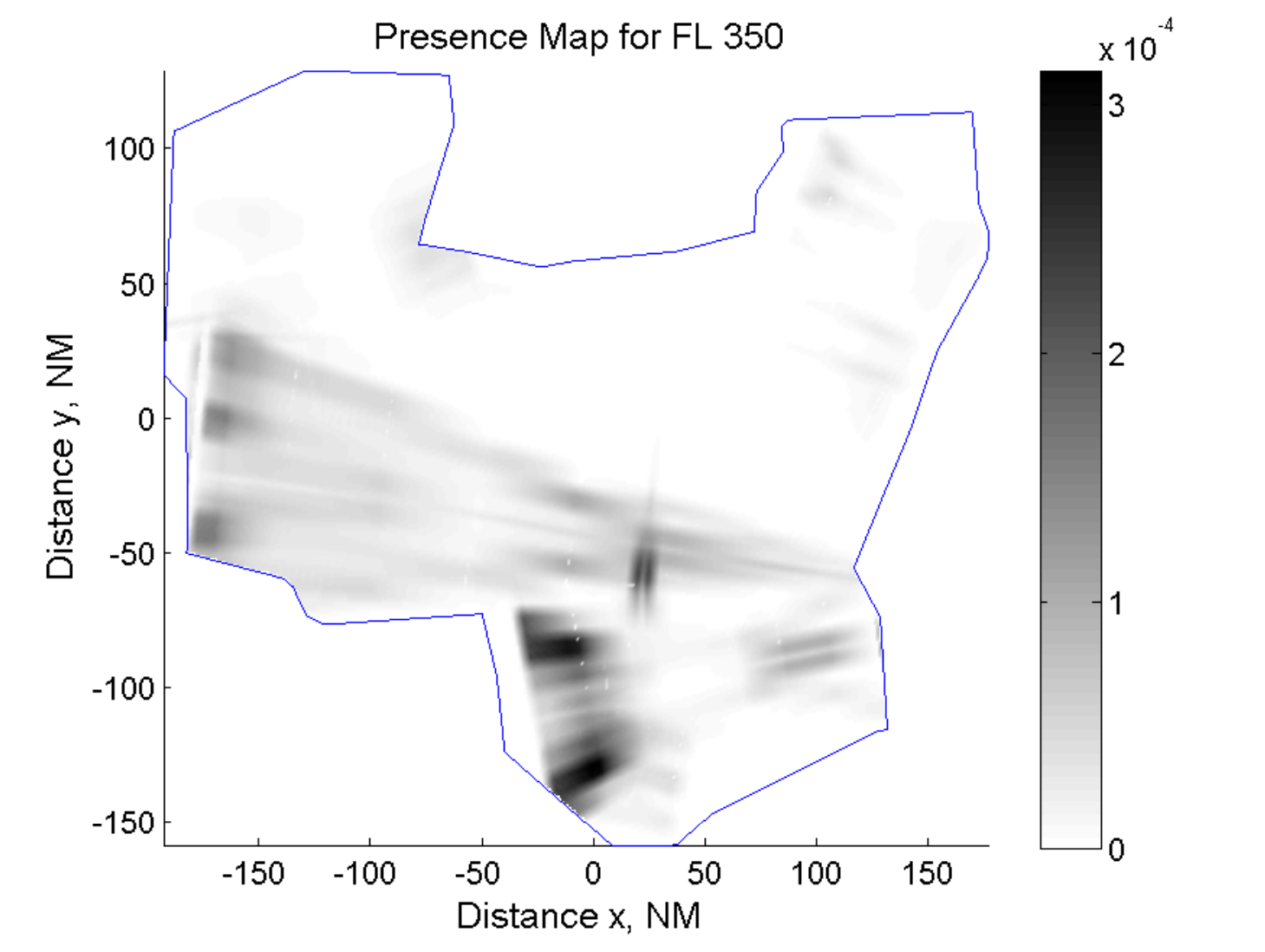}\label{first_map_3}}
\caption{Presence maps of Cleveland center ZOB between 2:00pm-2:15pm.}\label{first_map_total}
\end{figure}

\begin{figure}[htbp]
\centering
\subfigure[3D view]{\includegraphics[width=0.49\textwidth]{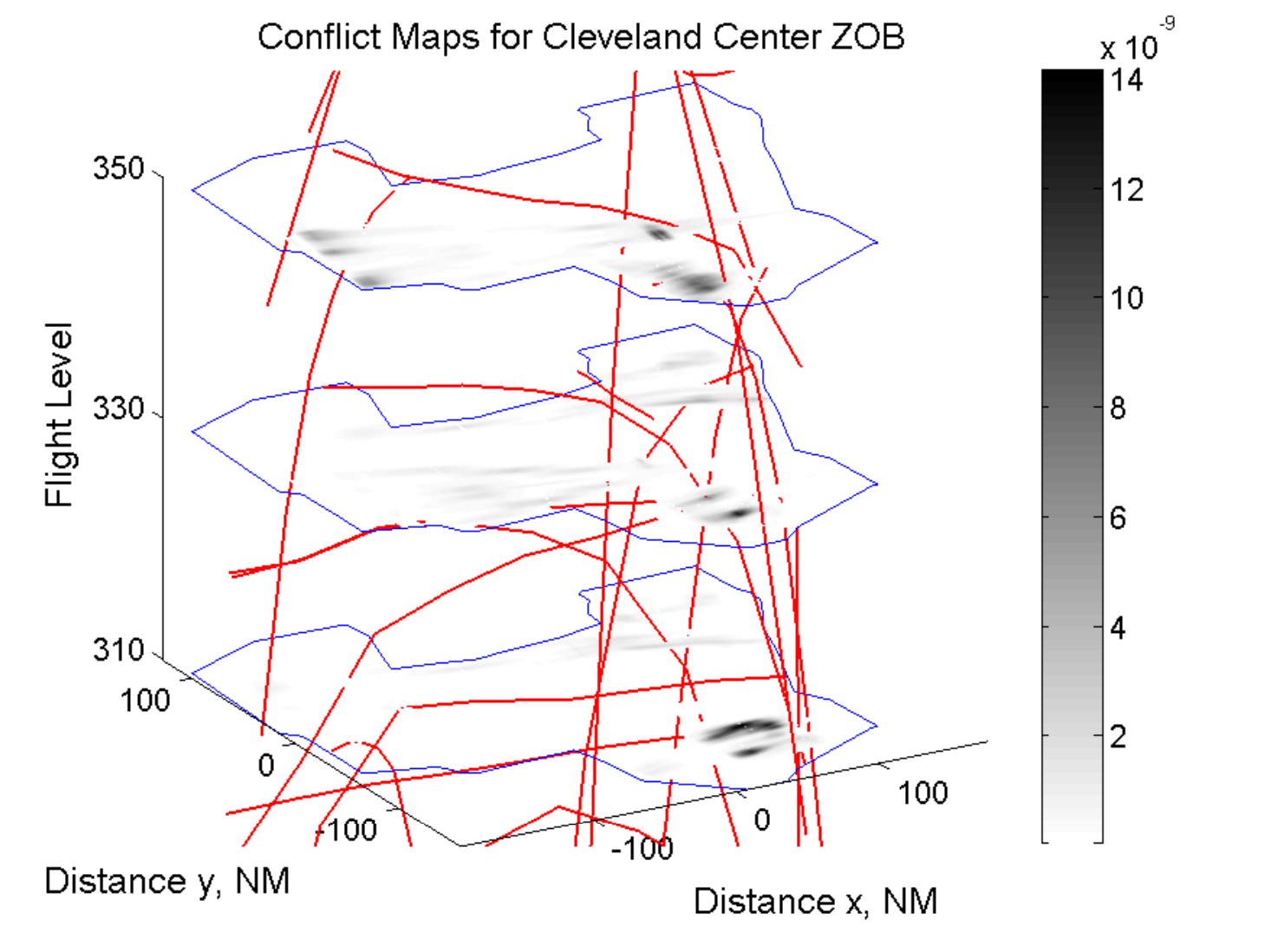}\label{second_map}}
\subfigure[2D view at FL 310]{\includegraphics[width=0.49\textwidth]{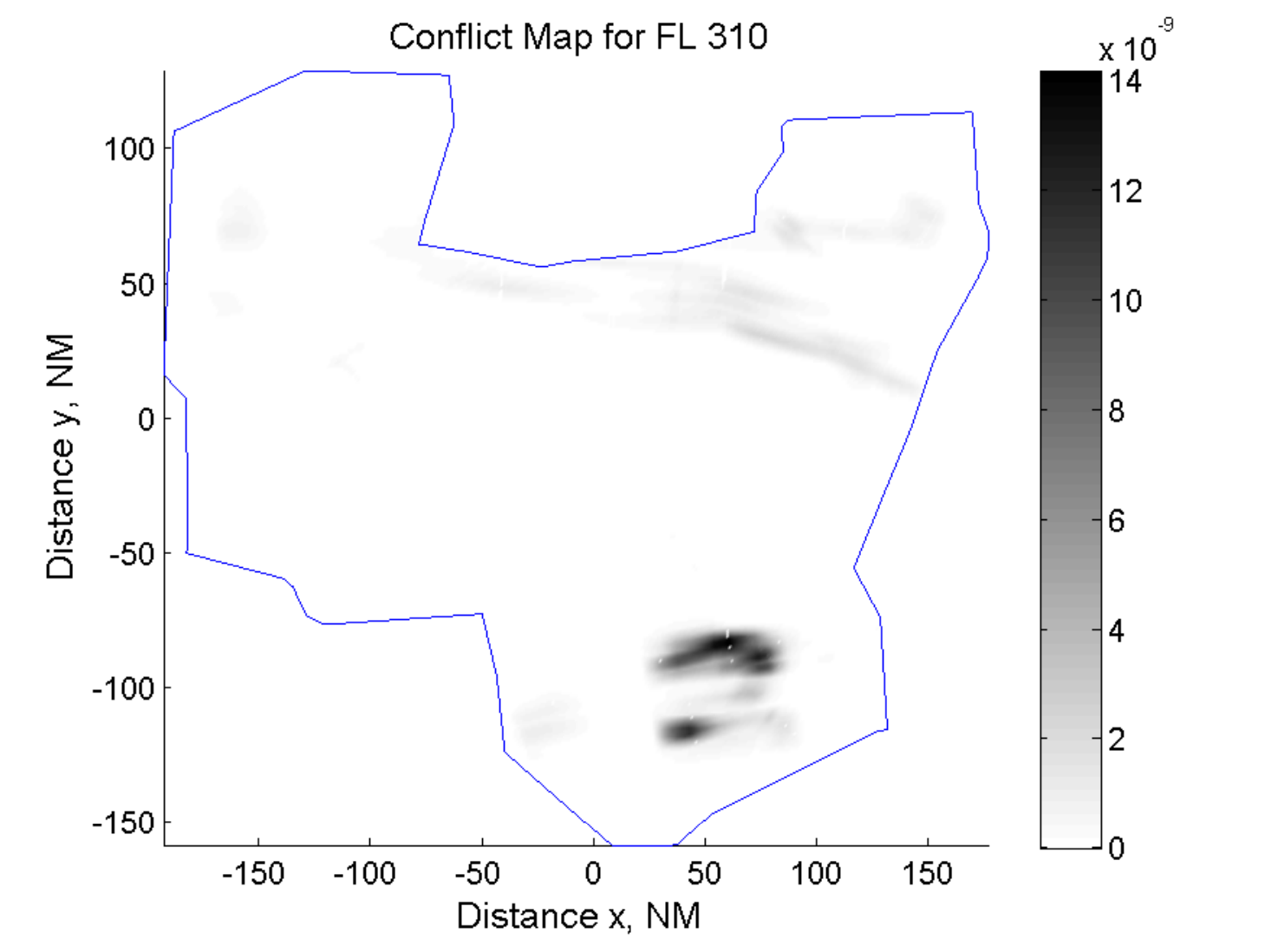}\label{second_map_1}}\\
\subfigure[2D view at FL 350]{\includegraphics[width=0.49\textwidth]{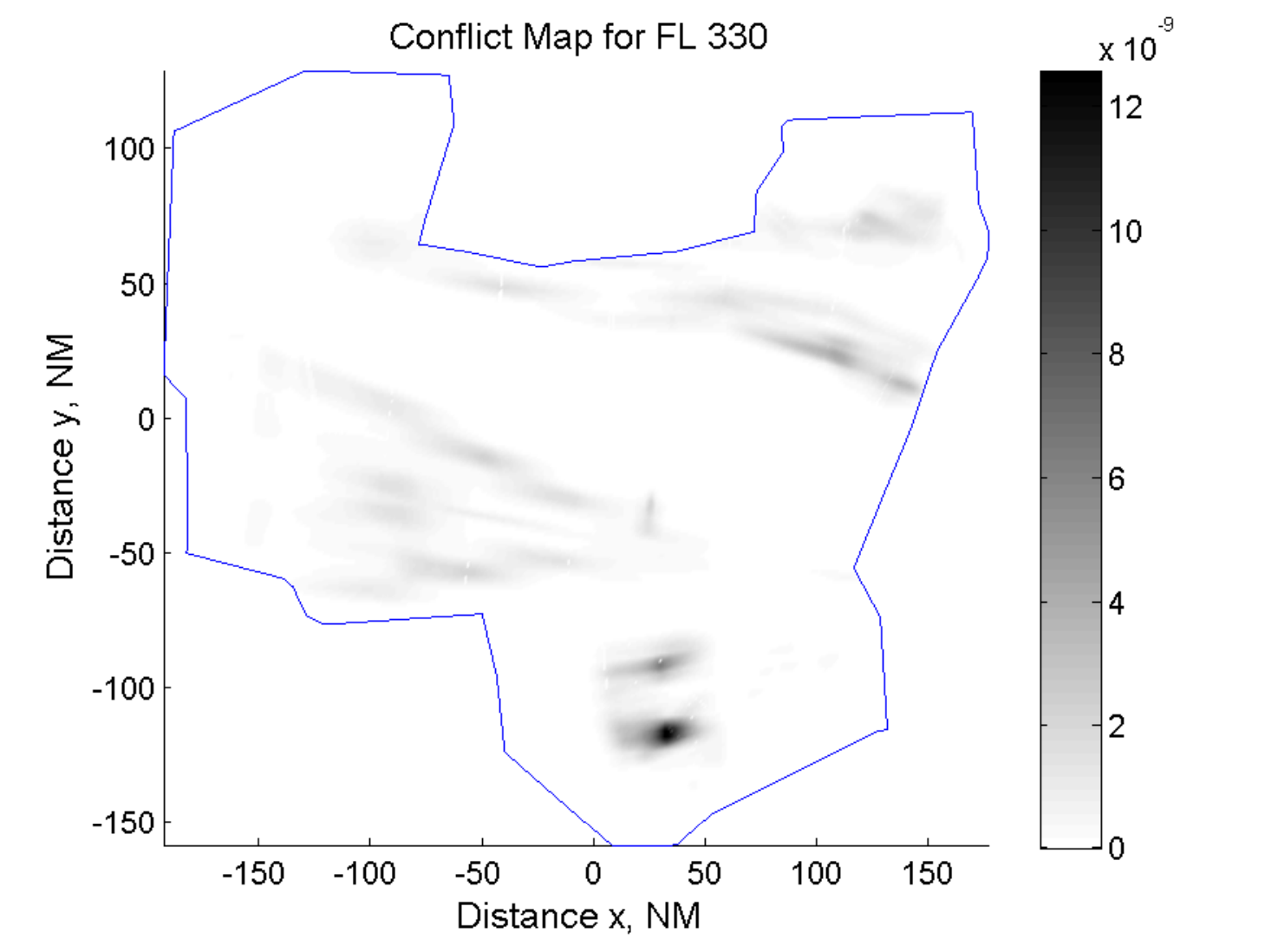}\label{second_map_2}}
\subfigure[2D view at FL 380]{\includegraphics[width=0.49\textwidth]{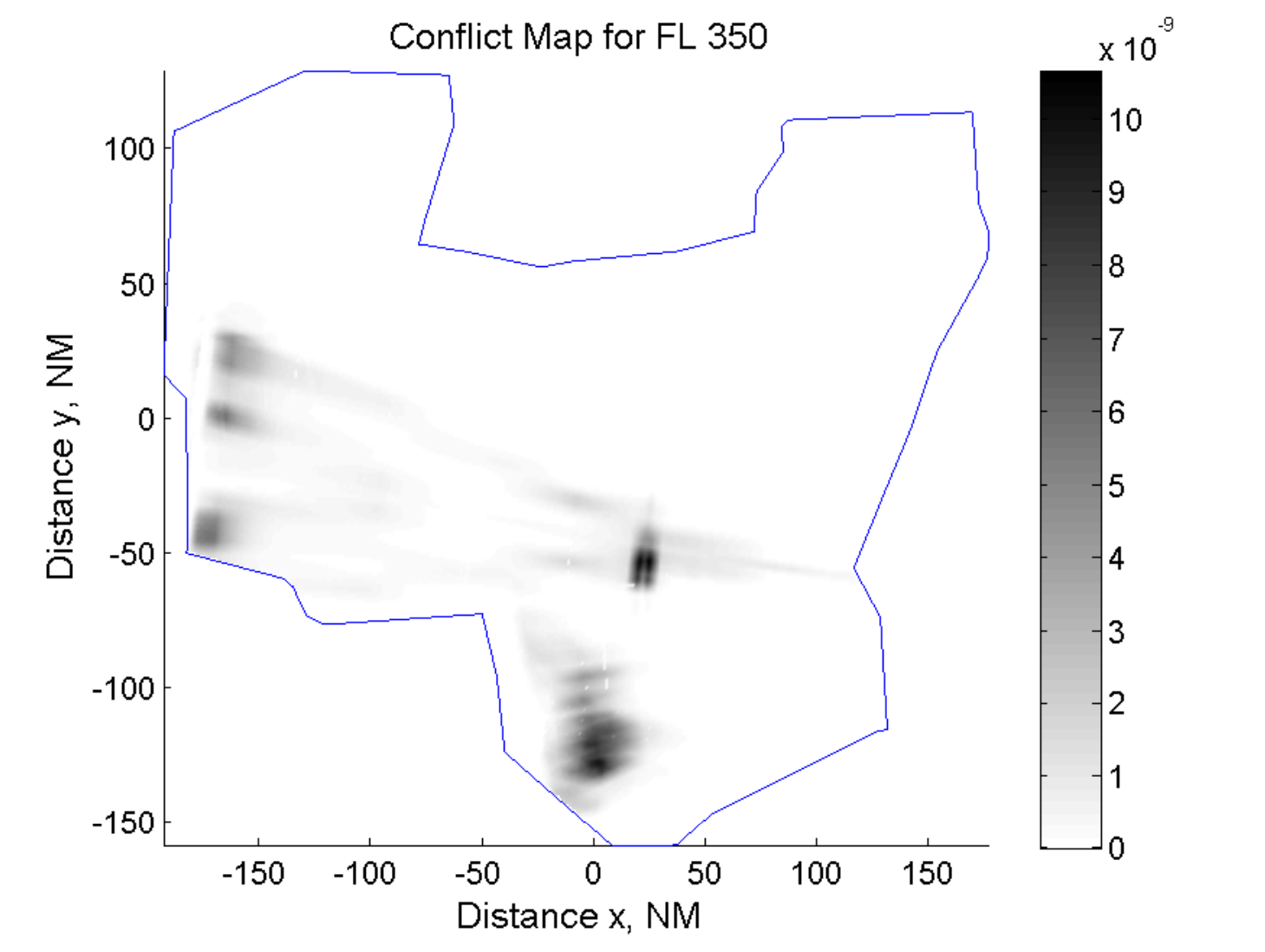}\label{second_map_3}}
\caption{Conflict maps of Cleveland center ZOB between 2:00pm-2:15pm.}\label{second_map_total}
\end{figure}

\begin{figure}[htbp]
\centering
\subfigure[2D view at FL 310]{\includegraphics[width=0.49\textwidth]{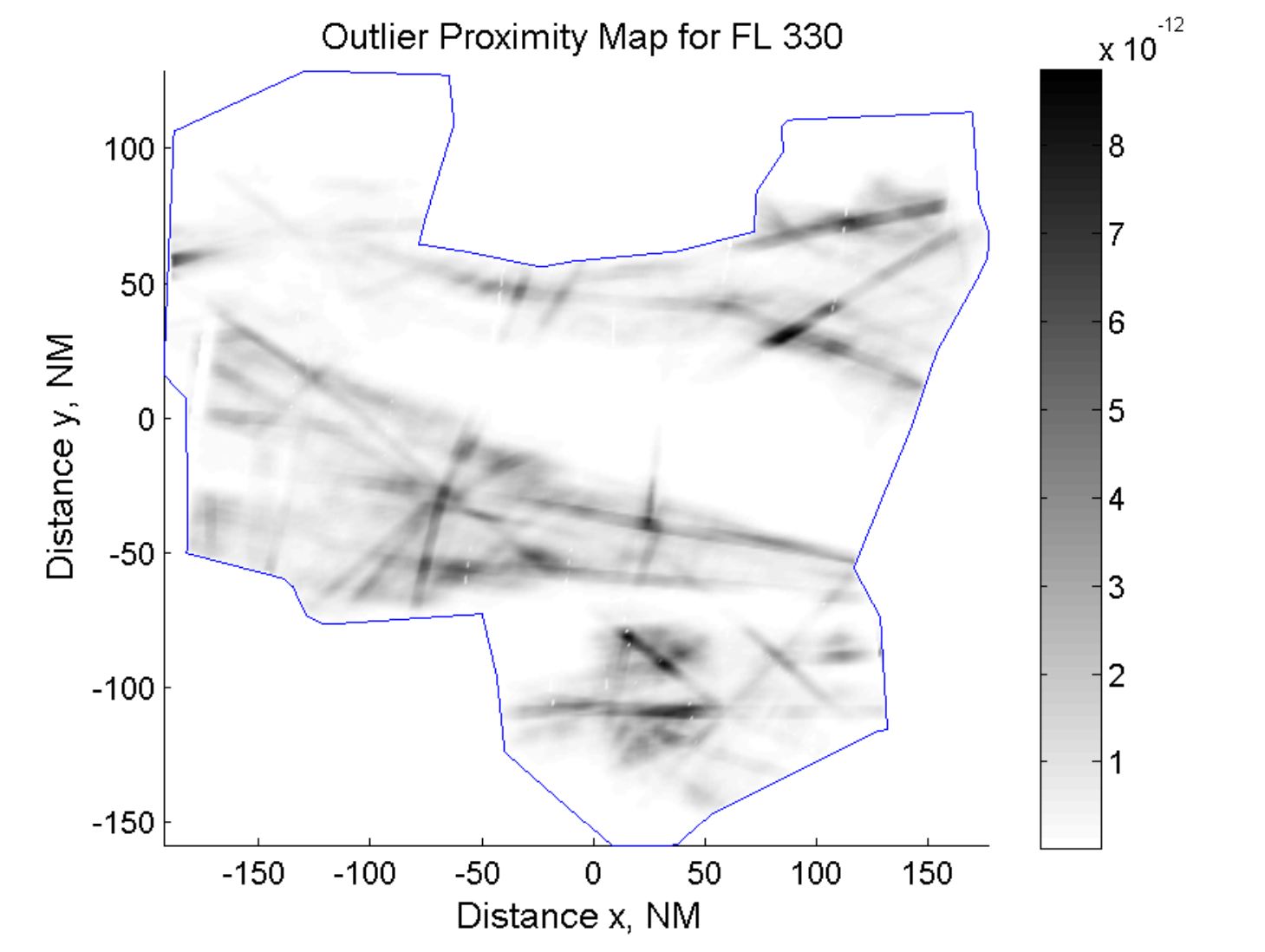}\label{outliers_map_1}}
\subfigure[2D view at FL 350]{\includegraphics[width=0.49\textwidth]{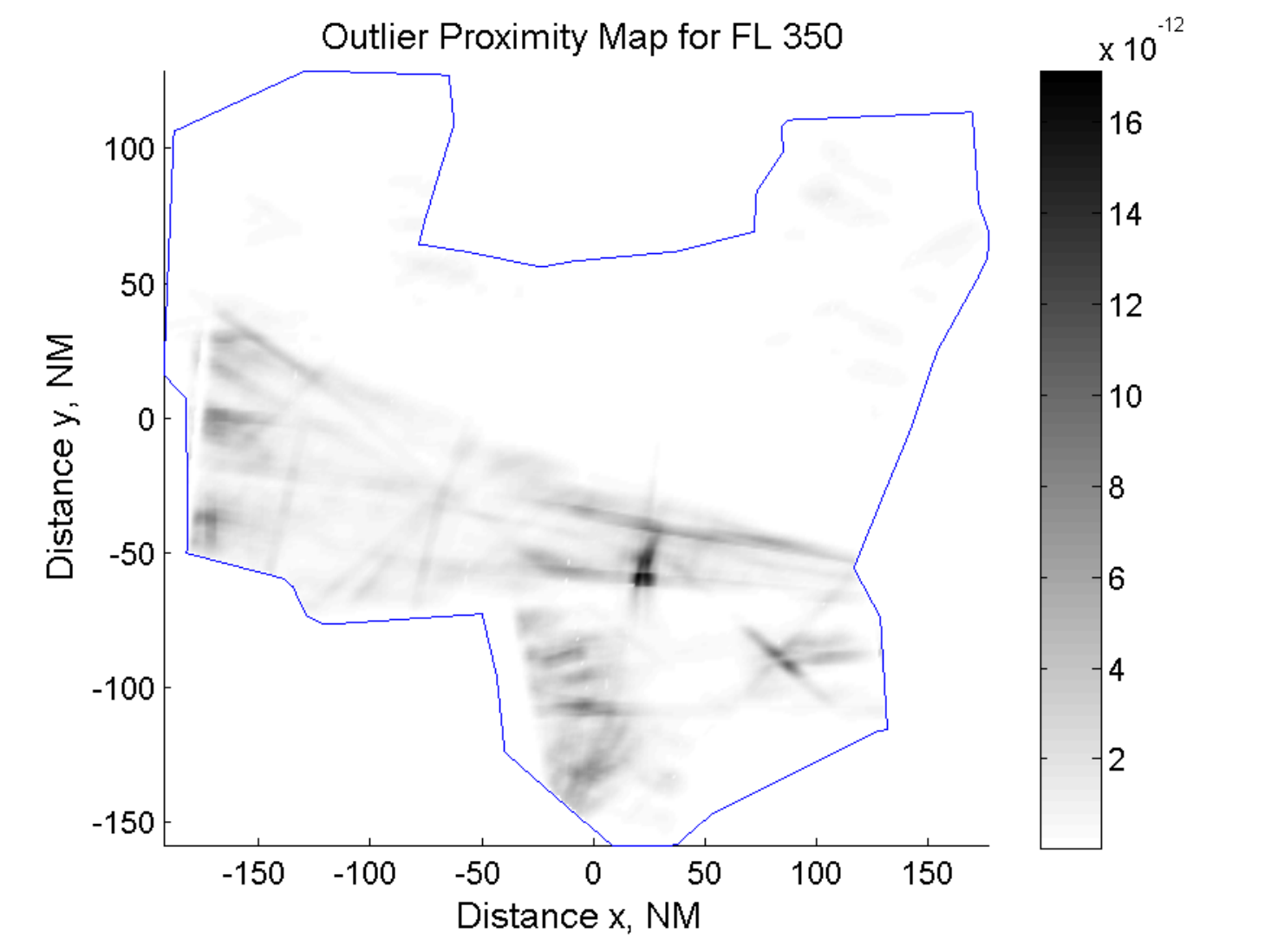}\label{outliers_map_2}}
\caption{Outlier proximity maps of Cleveland center ZOB between 2:00pm-2:15pm.}\label{outliers_map_total}
\end{figure}

This section combines both historical flow parameters determined from the clustering algorithm presented in Sections~\ref{sec:data} and~\ref{sec:gen_model}, and the methods to construct proximity maps given in Section~\ref{sec:proximity}.  The maps use the flows characteristics computed for May and the time period between 2:00p.m. and 2:15p.m. local time. During this time period, the expected number of entering aircraft is 65. For easier visualization of the results, proximity maps have been computed for the horizontal slice of airspace between FL~310 and FL~350. In this region, the mean inter-aircraft distances for the 10 busiest flows are given in Table~\ref{table_param}, and the centroids of the flows are plotted in red on Figs.~\ref{first_map} and~\ref{second_map}. Figures~\ref{first_map_total} and~\ref{second_map_total} illustrate the presence and conflicts maps, respectively. A 3D view and a 2D view are presented for each flight level. The outliers proximity maps for the considered center are given in Fig.~\ref{outliers_map_total}, which highlights the regions of the airspace that may be sensitive to the presence of an outlier. These maps provide an intuition of the importance of each flow, how they interact with each other, and how much influence an outlier can have on the main traffic. Since these maps provide a depiction of the probability of proximity of aircraft, they may be used in combination with a forecasted weather map, in order to estimate the regions where bad weather may have strong impact on the predicted traffic (as illustrated in~\cite{salaun:gnc10}). The maps can also be used as a tool for traffic flow managers to redesign routes or adjust flow rates in order to allow increased throughput while keeping a low level of proximity probability.

\begin{table}[htbp]
\centering
\begin{tabular}{|c|c|c|c|c|c|c|c|c|c|c|}
  \hline
   Flow number, $i$& 1 & 2 & 3 & 4 & 5 & 6 & 7 & 8 & 9 & 10 \\
  \hline
  $\Delta d_i^m$ (NM) & 45 & 56 & 57 & 88 & 90 & 101 & 105 & 116 & 140 & 225 \\
  \hline
\end{tabular}
\caption{Mean inter-aircraft distance, Cleveland center ZOB between 2:00pm-2:15pm}\label{table_param}
\end{table}

 \section{Conclusion}
\label{sec:conclusion}

This paper is concerned with computing airspace statistical complexity metrics using a parameterized generative model driven by historical flight data. Trajectories are clustered into major flows that account for 80\% of the traffic and the remaining 20\% are identified as outliers. These flows and outliers constitute a structure of air traffic, used as a parameterized generative model. Within the flows, aircraft spatial and temporal probability distributions are computed. Analysis of the airspace demonstrates that the aircraft arrival process into the center can be modeled as time-varying Poisson process. Using the structure determined by the flow and outliers model, three-dimensional maps are constructed to predict and visualize the probability of presence and probability of proximity of aircraft in the airspace.

The proximity maps are information displays that are relevant to traffic flow management.  Each map (presence, conflict, or outlier proximity map) identifies and highlights ``critical'' regions of the airspace that require significant monitoring.  Further, because the maps are based on parameterized historical traffic models, it is possible to consider ``what-if'' scenarios, in which parameters (flow rates and distributions) are adjusted in real-time to consider daily traffic variations.  Ultimately, this tool provides an information display to aid in the strategic and operational decision making required in traffic flow management.


\section*{Acknowledgements}
\noindent This work is funded by NASA under Grant NNX08AY52A and the FAA under Award No. 07-C-NE-GIT, Amendment Nos. 005, 010, and 020.






\end{document}